\documentclass[journal,comsoc]{IEEEtran}
\usepackage{cite}
\usepackage[pdftex]{graphicx}
\usepackage{caption}
\usepackage{epsfig}
\usepackage{epstopdf}
\usepackage{amsmath}
\usepackage{algorithmic}
\usepackage{algorithm}
\usepackage{amssymb}
\usepackage{empheq} 
\usepackage{color}
\usepackage{multirow}
\usepackage{multicol}
\usepackage{adjustbox}
\usepackage{graphicx}
\usepackage{multirow}
\usepackage[table,xcdraw]{xcolor}
\usepackage{subfig}
\usepackage{stfloats}
\usepackage{array}
\usepackage{longtable}
\usepackage{url}
\usepackage{eqnarray}
\usepackage{booktabs}
\usepackage{varwidth}
\usepackage{csquotes}
\usepackage[table]{xcolor}
\usepackage{pdflscape,array}
\usepackage{bm}
\usepackage{enumerate}
\usepackage{scalerel}
\usepackage{dsfont}
\pdfoutput=1

\begin{document}

\title{Energy Efficiency Maximization in IRS-enabled Phase Cooperative PS-SWIPT based Self-sustainable IoT Network}
\author{Haleema~Sadia,~Ahmad~Kamal~Hassan,~Ziaul~Haq~Abbas,~Ghulam~Abbas,~\textit{Senior Member,~IEEE},\\~Thar~Baker, \textit{Senior Member,~IEEE}
{\thanks{H. Sadia$^*$, A.K. Hassan, and Z.H. Abbas are with the Faculty of Electrical Engineering, Ghulam Ishaq Khan Institute of Engineering and Technology (GIKI), and  Telecommunications and Networking (TeleCoN) Research Center, Pakistan (e-mail: [haleema.sadia; akhassan; ziaul.h.abbas]@giki.edu.pk).}
\thanks{G. Abbas is with the Faculty of Computer Science and Engineering, GIKI, Pakistan (e-mail: abbasg@giki.edu.pk).}
\thanks{T. Baker is with the School of Architecture, Technology and Engineering, University of Brighton, Brighton BN2 4GJ, UK (e-mail: T.Shamsa@brighton.ac.uk).}}\vspace{-2ex}}


\maketitle
\begin{abstract}Power splitting based simultaneous wireless information and power transfer (PS-SWIPT) appears to be a promising solution to support future self-sustainable Internet of Things (SS-IoT) networks. However, the performance of these networks is constrained by radio frequency signal strength and channel impairments. To address this challenge, intelligent reflecting surfaces (IRSs) are introduced in PS-SWIPT based SS-IoT networks to improve network efficiency by controlling signal reflections. In this article, an IRS-enabled phase cooperative framework is proposed to improve energy efficiency (EE) of the IoT network $({\mathtt {I}}^{net})$ using phase shifts of the user network $({\mathtt {U}^{net})}$, without constraining hardware resources at ${\mathtt {U}^{net}}$. We exploit transmit beamforming (BF) at access points (APs) and phase shifts optimization at the IRS end with phase effective cooperation between APs to enhance ${\mathtt {I}}^{net}$ EE performance. The maximization problem turns out to be NP-hard, so first, an alternating optimization (AO) is solved for the ${\mathtt {U}^{net}}$ using low computational complexity heuristic BF approaches, namely, transmit minimum-mean-square-error and zero-forcing BF, and phase optimization is performed using semidefinite relaxation (SDR) approach. To combat the computational complexity of AO, we also propose an alternative solution by exploiting heuristic BF schemes and an iterative algorithm, i.e., the element-wise block-coordinate descent method for phase shifts optimization. Next, EE maximization is solved for the ${\mathtt {I}^{net}}$ by optimizing the PS ratio and active BF vectors by exploiting optimal phase shifts of the ${\mathtt {U}}^{net}$. Simulation results confirm that employing IRS phase cooperation in PS-SWIPT based SS-IoT networks can significantly improve EE performance of ${\mathtt {I}^{net}}$ without constraining resources.
\end{abstract}


 \section{Introduction}
 \label{sec:I}
\IEEEPARstart{W}{ith} upsurge growth in wireless devices and user-diverse application needs, significant contributions are being made to investigate beyond fifth-generation (B5G) networks. A plethora of these networks are projected to meet the anticipated demands of improved spectral efficiency (SE) and energy efficiency (EE) by providing ubiquitous, reliable, and near-instant services with low power consumption~\cite{9771335}. In transition to virtual reality, the next-generation Internet of Things (IoT) network appears as a paradigm shift and is envisioned to support a variety of wireless IoT devices (for example, smart phones, wearable devices, electronic tablets, smart security systems, sensors, etc.), paving the way for the future smart world~\cite{iannacci2023point}. Although these devices are equipped with diverse sensors, they are typically low-power and have limited battery life. Thus, these devices need to be maintained regularly in order to extend the battery's lifespan~\cite{liu2022wireless}. These constraints limit the potential use of IoT networks due to higher costs, deployment challenges, and increased network complexity~\cite{elsts2018enabling,vaezi2022cellular}.
\par
Recently, radio frequency (RF) transmission assisted simultaneous wireless information and power transfer (SWIPT) has received significant attention to address energy limitations and efficient information transfer for future self-sustainable (SS)-IoT networks~\cite{zhou2013wireless,8476597,8214104}.
SWIPT offers low power, usually in $\mu$W, but provides wide coverage in a sustainable and controllable manner. By exploiting power splitting-based SWIPT (PS-SWIPT), the PS-based devices in SS-IoT networks first perform energy harvesting (EH) and then information decoding (ID) by employing wireless power transfer (WPT) and wireless information transfer (WIT) techniques simultaneously~\cite{liu2016power,8951078,fang2023improper}.
Apart from PS-SWIPT based SS-IoT networks, many investigations have considered separate network architectures where WPT and WIT techniques are employed by different energy harvesting receivers (EHRs) and information decoding receivers (IDRs), respectively~\cite{9669263}.
PS-SWIPT based SS-IoT networks provide green communication with reduced capital expenditure as compared with separate network architectures~\cite{9680675}. However, due to different receiver sensitivities and application requirements in practical systems, these networks typically require significantly higher receive power in comparison to WIT networks~\cite{6805330}. Moreover, these networks are more susceptible to channel attenuation and are constrained by the received RF signal strength. Hence, in practical PS-SWIPT systems, poor efficiency of WPT for SS-IoT networks across large distances has been identified as the performance barrier. Although some studies have considered the massive multiple-input multiple-output (mMIMO) technique to enhance beamforming (BF) gain at the SWIPT transmitter to improve WPT efficiency, the main challenges for practical implementation remain high power consumption and hardware costs~\cite{9233408}.
\par
Intelligent reflecting surfaces (IRSs) have recently unlocked a novel communication paradigm to improve SE and EE for the sixth generation (6G) wireless networks~\cite{9690151,9895266}. It introduces the holographic idea of transmitting data by reusing an existing radio link, thus transforming the random wireless environment into a programmable one~\cite{9090356,9326394}. The low-cost passive reflective elements of IRS can be reconfigured intelligently to modify the phase reflections of the incident signal. IRS passive nature results in significantly lower power consumption without introducing any thermal noise as compared to conventional relay-aided communication systems that rely on active transmission sources~\cite{di2020reconfigurable}. IRS can be easily deployed for both indoor and outdoor applications because of their low cost and compact size. With these alluring features, IRS is already being deployed in conventional wireless networks to improve key performance metrics through the optimization of adaptive parameters such as active and passive BF, IRS phase shifts, transmission power, and effective IRS deployment~\cite{8811733,9771079}.
\par 
While the IRS serves to enhance conventional wireless communication networks performance, SWIPT applications are also attracted by the high BF gain achieved by passive IRS~\cite{9669263,li2021joint,10032536}. Many authors have considered IRS-supported WPT and WIT architectures with separate EHR and IDR to improve network performance. The authors in~\cite{8941080} investigated the optimization problem of weighted sum power maximization for EHRs. The authors demonstrated that the IRS-enabled SWIPT system did not require any specific energy-carrying signals and that sending information signals only to APs was adequate to service both IDR and EHR, independent of their channel realization. The authors in~\cite{9133435} solved transmit power minimization problem in a multi-IRS framework by jointly optimizing transmit precoders and IRS phase shifts, subject to the quality-of-service (QoS) requirements of all receivers, i.e., the individual signal-to-interference-plus-noise ratio (SINR) constraints at IDR and EH constraints for EHR. The authors in~\cite{9647934} proposed a new dynamic IRS-BF scheme to improve the throughput of IRS-enabled wireless communication framework. Joint optimization was performed to improve uplink and downlink wireless data transmission efficiency between hybrid access points and multiple wireless devices. The authors in~\cite{9110849} performed a joint optimization to maximize the weighted sum rate (WSR) of all IDRs subject to minimum EH requirements of all EHRs. An optimal resource allocation algorithm for the IRS-supported SWIPT system was proposed by the authors in~\cite{9669263}. The authors minimized the transmission power on the AP subject to the QoS requirements of IDR and nonlinear EHR by joint optimization of BF and the transmission mode selection strategy.
\par
To unveil the potential advantages of SS-IoT networks, some investigations have also considered the IRS-supported SWIPT framework with integrated IDR/EHR. The authors performed WSR maximization in~\cite{9680675} for a dynamic SS-IoT network based on MIMO PS-SWIPT with IRS support. The authors solved joint optimization problem with mutually exclusive constraints by providing an efficient optimization algorithm. The authors optimize the overall deployment of an IRS powered wireless sensor network (WPSN) within~\cite{9409104} by combining the optimized IRS phase shifts and the transmission time allocation. By exploiting IRS assistance in the proposed framework, the efficiency of WPT and WIT techniques were enhanced. The authors in~\cite{9423652} maximized the EE of the IRS-enabled PS-SWIPT system by considering both the information signal and the dedicated energy signal at the AP. For maximizing system throughput of wireless powered framework, the author in~\cite{9559408} provides maximization solution by optimizing the IRS phase shifts, transmission time, and bandwidth. In~\cite{10038636} the authors explored the same framework for multiple resource blocks with average throughput maximization as an objective. All of the aforementioned studies on IRS-enabled SWIPT IoT networks focused on maximization or minimization of different network metrics where IRS is deployed to serve IDRs and EHRs either separately or in integrated frameworks to improve the performance of next-generation IoT networks. However, to our knowledge, the problem of EE maximization in IRS-enabled phase cooperative PS-SWIPT based SS-IoT networks has not been explored yet.
 \subsection{Main Contributions}
  \label{sec:I-A}
Inspired by the potential applications of IRS-enabled SWIPT IoT networks, we propose an IRS-enabled phase cooperative framework to maximize the EE of a PS-SWIPT based SS-IoT network. The proposed framework comprises two distinct networks that operate under the frequency division multiple access (FDMA) scheme; the first network serves users equipment and is referred to as the $({\mathtt {U}^{net})}$, while the second network connects IoT devices and is referred to as the $({\mathtt {I}^{net})}$.
The IRS is primarily deployed for the ${\mathtt {U}^{net}}$; hence, the EE maximization problem is firstly formulated for this network and solved via an alternating optimization (AO) to get the optimal phase shifts. Later, the ${\mathtt {U}^{net}}$ IRS phase shifts are exploited to perform EE maximization at the ${\mathtt {I}^{net}}$ by applying effective phase cooperation. The proposed framework focuses on the use of IRS technology in SWIPT IoT networks, which can significantly reduce hardware costs.
The key contributions of this research are discussed next.
\begin{itemize}
\item This work presents an early attempt to investigate the EE maximization problem for IRS-enabled PS-SWIPT based SS-IoT networks using IRS phase cooperation. The optimization problem is developed to provide improved EE of the proposed framework by exploiting the phase cooperation between ${\mathtt{U}^{net}}$ and ${\mathtt{I}^{net}}$.
\item For the proposed framework, the EE maximization problem is solved by optimizing the transmit BF vectors of ${\mathtt{U}^{net}}$ and ${\mathtt{I}^{net}}$, and IRS phase shifts of ${\mathtt{U}^{net}}$. The formulated optimization problem turns out to be NP-hard and arduous to solve directly. Therefore, we first provide an AO solution by employing active transmit BF optimization and passive phase shifts optimization to solve optimization problem at the ${\mathtt{U}^{net}}$. Transmit beamformers are optimized by closed-form sub-optimal heuristic BF solutions, namely, minimum-mean-square-error (MMSE)/regularized zero-forcing-BF (ZFBF) and ZFBF. Further, to solve for IRS phase shifts optimization at the ${\mathtt{U}^{net}}$, a computationally efficient semidefinite relaxation (SDR) approach is employed. Moreover, the SDR relaxation's higher-rank solution is addressed by the Gaussian randomization method. 
\item Although the AO approach can solve the formulated problem efficiently, the SDR scheme results in significant computational complexity, particularly when a large number of IRS elements are used. To address this problem, we also propose an alternative low-complexity solution by leveraging the IRS's short-range/local coverage. This enables us to derive local optimal solutions to the phase shifts using a low-complexity iterative solution based on the element-wise block-coordinate descent (EBCD) method and BF vectors via a heuristic solution in closed-form. The proposed solutions' computational complexity and convergence analysis unveil potential implications for practical applications.
\item The performance gain in EE for the PS-SWIPT based SS-IoT network is assessed by optimizing power splitting coefficient, transmit BF using heuristic approach, and employing the ${\mathtt {U}^{net}}$ phase shifts via phase cooperation. Simulations results demonstrate that the EE of PS-SWIPT based SS-IoT networks can be significantly improved by employing IRS phase cooperation which significantly reduces hardware installation cost without constraining resources at ${\mathtt {U}^{net}}$.
\end{itemize}
\subsection{Paper Structure and Notations}
 \label{sec:I-B}
We have organised this article into seven sections. Following an introduction in Section~\ref{sec:I}, Section~\ref{sec:II}, presents the proposed framework and related optimization problem formulations. Section~\ref{sec:III} describes the AO technique for optimizing transmit beamformers and phase shifts. Section~\ref{sec:IV} discusses the low complexity alternative solution for the given problem. The EE optimization problem with the proposed solutions for IoT network is discussed in Section~\ref{sec:V} of the paper. Simulation findings and related discussion is provided in Section~\ref{sec:VI}. Section~\ref{sec:VII} outlines our research conclusion and future directions.
\par
\emph{Notations}: In this work, scalar, vector, and matrix are represented by bold, lowercase, and uppercase letters, respectively. The complex matrix of the space $a\times b$ is denoted by the symbol $\mathbb {C}^{a\times b}$ and $\mathcal {CN}$ represents the complex Gaussian random variable. The absolute square, Euclidean norm, and Hermitian operator of a complex-valued vector $\textbf a$ are denoted by $|a|^2$, $||\textbf A||^2$, and $(\textbf a)^H$, respectively. The summation operator, real part of a complex value, and statistical expectation of random variables are represented by $\sum(\cdot)$, $\mathfrak{R}[\cdot]$, and $\mathbb{E}[\cdot]$, respectively.

\section{System Model and Problem Formulation}
\label{sec:II}
 \setlength\belowcaptionskip{-2ex}
\begin{figure}
\centering
\includegraphics[width=3.4in]{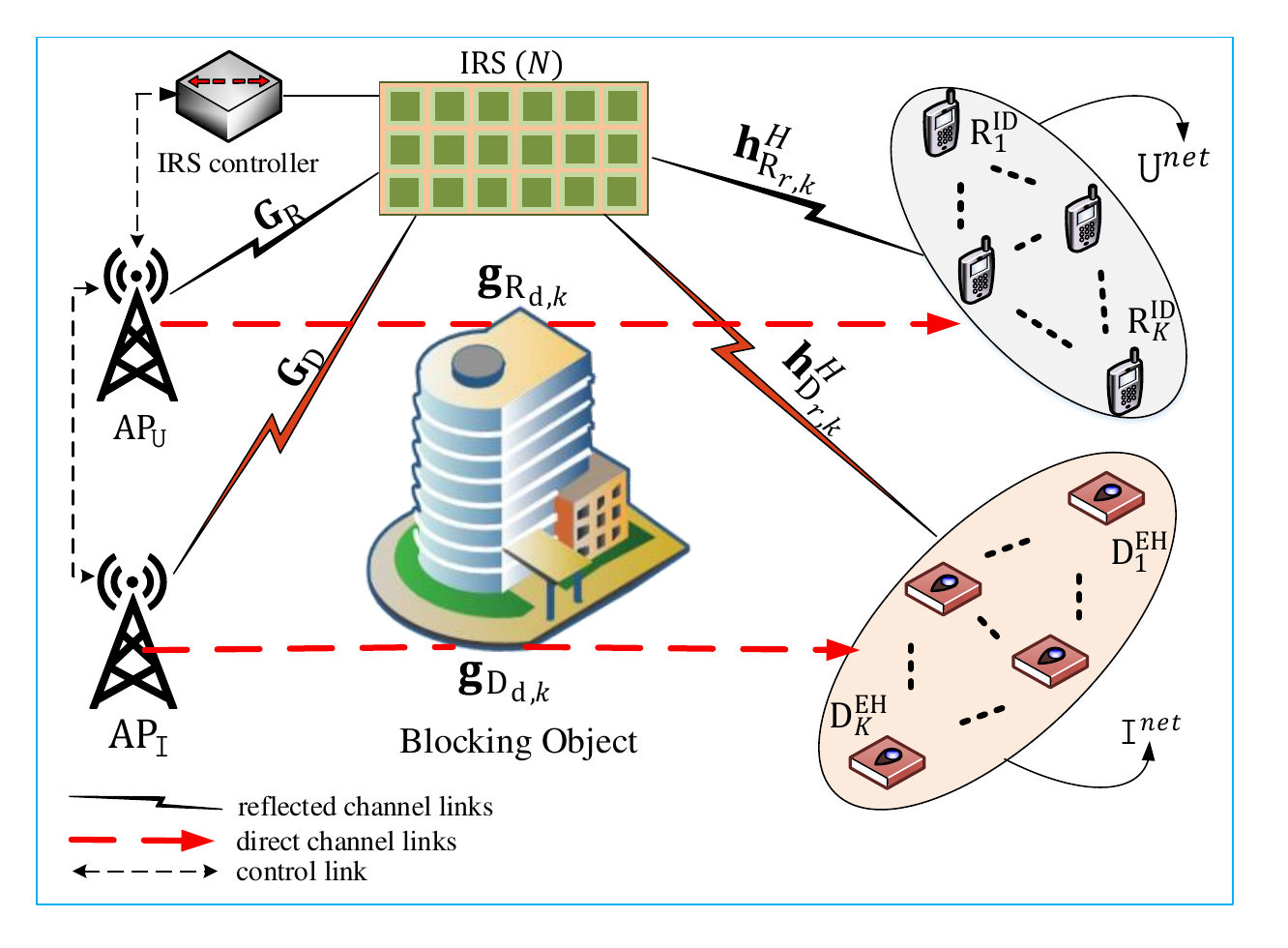}\caption{\small {IRS-aided phase cooperative framework with users and PS-SWIPT based SS-IoT networks.}}
\label{fig:1}
\end{figure}
\subsection{IRS-aided Phase Cooperative Network}
 \label{sec:II-A}
Fig.~\ref{fig:1} shows an IRS-enabled downlink phase cooperative framework with a pair of ${\mathtt {U}^{net}}$ having $K_{\rm I}$ information decoding receivers/equipment's $({{{\rm R}_k}})$ and ${\mathtt {I}}^{net}$ having $K_{\rm E+I}$ EH and information decoding devices $({{\rm D}_k})$. Both the networks are served through separate access points $(\rm APs)$, i.e., $({\rm {AP}_\mathtt {U}})$ and $({\rm {AP}_\mathtt {I}})$ for information transfer and energy$+$information transfer to ${\mathtt {U}^{net}}$ and ${\mathtt {I}^{net}}$, respectively under multiple-input single-output (MISO) network configuration. The ${\rm {AP}_\mathtt {U}}$ is fitted with $M_\mathtt {U}$ and ${\rm {AP}_\mathtt {I}}$ with $M_\mathtt {I}$ transmit antennas such that $M_\mathtt {U} \gg{K_{\rm I}}$ and $M_\mathtt {I}\gg{K_{\rm E+I}}$, while each receiver pair has a single antenna. A frequency division multiple access (FDMA) protocol is predicated between APs for the channel access, in which both ${\mathtt {U}^{net}}$ and ${\mathtt {I}^{net}}$ pairs operate on distinct license spectrum. Further, within each network, users access channel via space-division multiple access (SDMA) scheme, where $K_{\{{\rm I, E+I}\}}$ spatially separated receivers and devices are served simultaneously. An IRS with ${N}$ low-cost passive reflection elements equipped with a smart controller is installed between the ${\rm {AP}_\mathtt {U}}$ and ${{{\rm R}_k}}$. The ${\rm {AP}_\mathtt {I}}$ shares the ${\mathtt {U}^{net}}$'s IRS according to phase cooperation by exploiting the optimal phase shifts. The IRS controller maintains a separate wireless link with the APs to communicate channel state information (CSI), transmission data, and to receive IRS reflection coefficients designed at ${\rm {AP}_\mathtt {U}}$. Furthermore, for effective phase shifts cooperation a dedicated link is used between ${\rm {AP}_\mathtt {U}}$ and ${\rm {AP}_\mathtt {I}}$. Moreover, we assumed perfect CSI at the APs\footnote{Although the CSI of IRS-enabled networks is often challenging to obtain, this work focuses on the upper bound for the system under investigation~\cite{9707728}. As a result, we presume perfect CSI for both APs and IRS~\cite{ding2020simple,ding2020impact}. However, the accurate CSI can be obtained via multiple channel estimation algorithms investigated in the literature for IRS-enabled systems~\cite{he2019cascaded,wang2020channel,taha2021enabling,10053657}. Additionally, the proposed framework can be exploited further with an imperfect CSI scenario~\cite{wang2022safeguarding}.}, where information transfer between all nodes is solely performed at APs. The users are located in the vicinity of blocking objects; hence, they receive both IRS reflected (r) and direct (d) signals.
\par 
Now, consider that ${\rm {AP}}_\mathtt {U}$ and ${\rm {AP}_\mathtt {I}}$ broadcast $K$ different messages spatially separated for transfer of information and energy signals to ${\mathtt {U}^{net}}$ and ${\mathtt {I}^{net}}$, i.e., ${s_i}$ $\{i \in K_{\rm I}\}$ and ${s_j}$ $\{j \in K_{\rm E+I}\}$ with normalized power as $\mathbb{E}\{| {s_i}|^2\}=1$ and $\mathbb{E}\{|{s_j}|^2\}=1$. At ${\mathtt {U}^{net}}$ the signal received by ${{{\rm R}_k}}$ with linear active transmit BF vector $\textbf{w}_{{\rm R}_{k}}\in\mathbb{C}^{M_\mathtt {U}\times 1}$ at ${\rm {AP}_\mathtt {U}}$ and passive phase shifts $(\Theta)$ at IRS after $r$ multi-path reflections is given as in eq.~\eqref{eq:1}
\setlength{\abovedisplayskip}{2pt}
\setlength{\belowdisplayskip}{2pt}
\begin{align}
y_{{\rm R}_{k}} &=(\textbf{h}_{{{{\rm R}_{{\rm r},k}}}}^H \Theta \textbf{G}_{\rm R}+\textbf{g}_{{{{\rm R}_{{\rm d},k}}}}^H)\displaystyle\sum_{i=1}^{K_{\rm I}}
{{{\textbf w}_{{\rm R}_{i}}{s_i}}}+{z_{{\rm R}_{k}}},\ \forall k \in {K_{\rm I}}.
\label{eq:1}
\end{align}
Next, ${\rm {AP}_{\mathtt {I}}}$ uses ${\mathtt {U}^{net}}$'s IRS phase shifts $\Theta$, therefore at ${\mathtt {I}^{net}}$ the signal received by the ${{\rm D}_k}$ using ${\rm {AP}_\mathtt {I}}$ active transmit BF vector $\textbf{w}_{{\rm D}_{k}}\in\mathbb{C}^{M_\mathtt {I}\times 1}$ is given in eq.~\eqref{eq:2} 
\begin{align}
y_{{\rm D}_{k}} &=(\textbf{h}_{{{{\rm D}_{{\rm r},k}}}}^H \Theta \textbf{G}_{\rm D}+\textbf{g}_{{{{\rm D}_{{\rm d},k}}}}^H)\displaystyle\sum_{j=1}^{K_{\rm E+I}}{{{\textbf w}_{{\rm D}_{j}}{s_j}}}+z_{{\rm D}_{k}},\ \forall k \in {K_{\rm E+I}},
\label{eq:2}
\end{align}
where in~(\ref{eq:1}) and~(\ref{eq:2}), $\textbf{h}_{{{{\rm R}_{{\rm r},k}}}}^H \in\mathbb{C}^{1\times N}$,  $\textbf{h}_{{{{\rm D}_{{\rm r},k}}}}^H\in\mathbb{C}^{1 \times N}$ are the channel gains from the IRS to the ${\mathtt {U}^{net}}$ and ${\mathtt {I}^{net}}$, respectively and  $\textbf{G}_{\rm R}\in\mathbb{C}^{N \times M_\mathtt {U}}$, $\textbf{G}_{\rm D}\in\mathbb{C}^{N \times M_\mathtt {I}}$ show the respective channel gains from ${\rm AP}_\mathtt {U}$ and ${\rm AP}_\mathtt {I }$ to the IRS. The direct channel links from APs to users and IoT devices are represented by $\textbf{g}_{{{{\rm R}_{{\rm d},k}}}}^H$ and $\textbf{g}_{{{{\rm D}_{{\rm d},k}}}}^H$, respectively. The passive reflection matrix of ${\mathtt {U}^{net}}$, represented by the diagonal matrix $\Theta \in\mathbb{C}^{N\times N}$, provides information on the amplitude and the phase shifts reflection coefficients, i.e.,
\setlength{\abovedisplayskip}{2pt}
\setlength{\belowdisplayskip}{2pt}
 \begin{equation}
\Theta={\rm diag}(\beta_{{{1}}}{e^{j\theta_1}},\beta_{{{2}}}{e^{j\theta_2}}\ldots,\beta_{N}e{^{j\theta_{N}}}),
 \label{eq:3}
 \end{equation} 
where $\beta_{{1}},\ldots,\beta_{N}\in[0,1]$ and $\theta_1,\ldots,\theta_{N}\in{[0,2\pi)}$ show the fixed amplitude reflection coefficients and IRS phase shifts variables, respectively. In literature, the amplitude reflection model is frequently assumed fixed, i.e., $\beta_{n}=1, \forall n \in N$ to eliminate hardware complexity overhead~\cite{9326394}. 
The factors ${ {z_{{\rm R}_{k}}}\in\mathcal{CN}(0,\sigma_{k}^2)}$, and ${ {z_{{\rm D}_{k}}}\in\mathcal{CN}(0,\sigma_{k}^2)}$ denote the respective receiver additive noise at ${{{\rm R}_k}}$ and ${{\rm D}_k}$.
\par
At ${\mathtt {U}^{net}}$, the SINR expression for the $k{\rm th}$ receiver is given by eq.~\eqref{eq:4}
\begin{equation}
\Gamma_{{\rm R}_{k}}=\frac{|(\textbf{h}_{{{\rm R}_{{\rm r},k}}}^H \Theta \textbf{G}_{\rm R}+\textbf{g}_{{{\rm R}_{{\rm d},k}}}^H)\textbf{w}_{{\rm R}_{k}}|^2}{\displaystyle\sum_{i\neq k}{|(\textbf{h}_{{{\rm R}_{{\rm r},k}}}^H \Theta \textbf{G}_{\rm R}+\textbf{g}_{{{\rm R}_{{\rm d},k}}}^H)\textbf{w}_{{\rm R}_{i}}|^2+\sigma_{k}^2}}, \ \forall k \in K_{\rm I},
\label{eq:4}
\end{equation}
where ${\sum_{i\neq k}{|(\textbf{h}_{{{\rm R}_{{\rm r},k}}}^H \Theta \textbf{G}_{\rm R}+\textbf{g}_{{{\rm R}_{{\rm d},k}}}^H)\textbf{w}_{{\rm R}_{i}}|^2}}$ is representing interference term encountered by $k{\rm th}$ receiver.
\par
The achievable rate of ${{\rm R}_{k}}$ can be denoted in eq.~\eqref{eq:5} by
\begin{align}
    {\mathcal R}_{{\rm R}_{k}}&=\ {\rm log}_2 (1+\Gamma_{{\rm R}_{k}}), \ \forall k \in K_{\rm I}.
     \label{eq:5}
\end{align}
\par
Accordingly, the sum rate of ${{\mathtt {U}^{net}}}$ is calculated as
\begin{align}
   {\mathcal R}_{k}^{{\mathtt {U}^{net}}}&={{\displaystyle\sum_{k=1}^{K_{\rm I}}}{\mathcal R}_{{\rm R}_{k}}}, \ \forall k \in K_{\rm I}.
    \label{eq:6}
\end{align}
\subsection{PS-SWIPT Implementation at SS-IoT Network}
 \label{sec:II-B}
In order to decode information signals, the SS-IoT devices require adequate energy, which is usually not available because of the low energy resources of the IoT network. Therefore, the received signal power is split for EH and ID by PS-based SS-IoT devices~\cite{xu2019robust,8879665}. Power splitting enables the simultaneous implementation of EH and ID ~\cite{liu2013wireless}, in contrast to time switching, which mandates distinct time slots for both processes~\cite{nasir2015wireless}. In order to efficiently decode their data, the devices used part of the received signal power for EH and the remaining part for ID. We assume that the received signal power is divided into $({\varphi_k})$ and $(1-{\varphi_k})$ as the power harvesting coefficient and the ID coefficient by ${{\rm D}_{K_{\rm I } }}$ devices, respectively. Furthermore, as we are employing a power splitting scheme, we consider that the entire transmission is taking place in a single time slot, $T$.
\par
IoT devices receive a fraction of the power from $y_{{\rm {D}_k}}$, that is, ${\varphi_k}$ for EH and the remaining fraction $(1-{\varphi_k})$ for decoding information.
After performing EH, the received signal available for ID is represented as in eq.~\eqref{eq:7}
\begin{align}
y_{{\rm D}_k}^{\rm {ID}} &=\sqrt{1-{\varphi_k}}y_{\rm {D}_k}+{\rm \omega_{\rm eh}}\notag
\\
&=\sqrt{1-{\varphi_k}}(\textbf{h}_{{{{\rm D}_{{\rm r},k}}}}^H \Theta \textbf{G}_{\rm D}+\textbf{g}_{{{{\rm D}_{{\rm d},k}}}}^H)\displaystyle\sum_{j=1}^{K_{\rm E+I}}{{{\textbf w}_{{\rm D}_{j}}{s_j}}}+\notag \\ 
& \ \ \ \ \sqrt{1-{\varphi_k}}z_{{\rm D}_{k}} + {\rm \omega_{\rm eh}}, \ \forall k \in {K_{\rm E+I}},
\label{eq:7}
\end{align}
here, ${\rm \omega_{\rm eh}}$ is representing the thermal noise introduced by the circuitry due to phase offset (with zero mean and $\sigma^2$ variance)~\cite{7934322}. Furthermore, the noise factor after EH, i.e., $(\sqrt{1-{\varphi_k}})z_{{\rm D}_{k}}$ is very small and can be ignored. Consequently, the information signal received at $k{\rm th}$ SS-IoT device (${{\rm D}_k}$) after EH is represented as in eq.~\eqref{eq:8}
\begin{equation}
y_{{\rm D}_k}^{\rm ID} =(\sqrt{1-{\varphi_k}})(\textbf{h}_{{{{\rm D}_{{\rm r},k}}}}^H \Theta \textbf{G}_{\rm D}+\textbf{g}_{{{{\rm D}_{{\rm d},k}}}}^H)\displaystyle\sum_{j=1}^{K_{\rm E+I}}{{{\textbf w}_{{\rm D}_{j}}{s_j}}}+ {\rm \omega_{\rm eh}},\ \forall k,
\label{eq:8}
\end{equation}
The corresponding SINR expression is given by
\begin{align}
\Gamma_{{\rm D}_{k}}^{\rm ID}&=\frac{{(1-{\varphi_k})}{|(\textbf{h}_{{{\rm D}_{}\rm r},k}^H \Theta \textbf{G}_{\rm D}+\textbf{g}_{{{\rm D}_{}\rm d},k}^H)\textbf{w}_{{\rm D}_{k}}|^2}}{\displaystyle\sum_{j\neq k}{(1-{\varphi_k})}{|(\textbf{h}_{{{\rm D}_{{\rm r},k}}}^H \Theta \textbf{G}_{\rm D}+\textbf{g}_{{{\rm D}_{}\rm d},k}^H)\textbf{w}_{{\rm D}_{j}}|^2+\sigma_{k}^2}},\ \forall k,
\label{eq:9}
\end{align}
where ${\sum_{j\neq k}{(1-{\varphi_k})}{|(\textbf{h}_{{{\rm D}_{{\rm r},k}}}^H \Theta \textbf{G}_{\rm D}+\textbf{g}_{{{\rm D}_{}\rm d},k}^H)\textbf{w}_{{\rm D}_{j}}|^2}}$ denotes the interference term encountered by the $k{\rm th}$ device of ${\mathtt {I}^{net}}$. Consequently, the achievable rate of ${{\rm R}_{k}}$ and the sum rate of ${{\mathtt {I}^{net}}}$ can be stated in eq.~\eqref{eq:10} and eq.~\eqref{eq:11}, respectively as
\begin{align}
    {\mathcal R}_{{\rm D}_{k}}&=\ {\rm log}_2 (1+\Gamma_{{\rm D}_{k}}^{\rm ID}), \ \forall k \in K_{\rm E+I}.
    \label{eq:10} \\
  {\mathcal R}_{k}^{{\mathtt {I}^{net}}}&={{\displaystyle\sum_{k=1}^{K_{\rm E+I}}}{\mathcal R}_{{\rm D}_{k}}}, \ \forall k \in K_{\rm E+I}.
    \label{eq:11}
\end{align}
\par
Similarly, the received signal for EH is given by eq.~\eqref{eq:12}
\begin{align}
y_{{\rm D}_k}^{\rm {EH}} &=\sqrt{{\varphi_k}}y_{\rm {D}_k}\notag \\
&=\sqrt{{\varphi_k}}(\textbf{h}_{{{{\rm D}_{{\rm r},k}}}}^H \Theta \textbf{G}_{\rm D}+\textbf{g}_{{{{\rm D}_{{\rm d},k}}}}^H)\displaystyle\sum_{j=1}^{K_{\rm E+I}}{{{\textbf w}_{{\rm D}_{j}}{s_j}}}+\sqrt{{\varphi_k}}z_{{\rm D}_{k}},\ \forall k,
\label{eq:12}
\end{align}
here, the power harvested from additive noise, i.e., $\sqrt{{\varphi_k}}z_{{\rm D}_{k}}$, is very low compared with signal harvesting power and hence can be neglected~\cite{9417357,9423652}. The ${{\rm D}_{k}}$ can only decode their data when they have a sufficient amount of energy. Therefore, the amount of energy harvested by ${{\rm D}_{k}}$ is provided by eq.~\eqref{eq:13}
\begin{align}
    E_{\mathcal H_k}={\varphi_k}\displaystyle\sum_{k=1}^{K_{\rm E+I}}{|(\textbf{h}_{{{\rm D}_{{\rm r},k}}}^H \Theta \textbf{G}_{\rm D}+\textbf{g}_{{{\rm D}_{{\rm d},k}}}^H)\textbf{w}_{{\rm D}_{k}}|^2}\mu_k,\ \forall k \in K_{\rm E+I},
\label{eq:13}
\end{align}
where ${\mu_k} \in [0,1]$ represents the circuitry's EH efficiency. The PS-SWIPT technique is utilized effectively at ${\mathtt {I}^{net}}$, which results in effective communication by exploiting passive IRS phase shifts at ${\mathtt {U}^{net}}$ and active BF at ${\rm {AP}_\mathtt {I}}$. 

\subsection{Formulation of Optimization Problem}
 \label{sec:II-C}
For the proposed IRS-enabled ${{\rm PC}_{Net}}$ framework, our objective is to maximize the EE of ${\mathtt {I}^{net}}$. Therefore, to perform EE maximization, we first optimize the power harvesting coefficient, which improves ${\mathtt {I}^{net}}$ EH power. Next, we optimize the precoding vectors ${{\textbf w}_{{\rm D}_{k}}}$ of ${\rm {AP}_\mathtt {I}}$ by exploiting the optimal phase shifts of ${\mathtt {U}^{net}}$ while keeping the transmit power of ${\rm {AP}_\mathtt {I}}$ to a minimum. For solving optimization problems, we initially calculate the total power consumption in the proposed framework, which depends on multiple parameters. Therefore, for ${\mathtt {I}^{net}}$, let us define $\eta \in [0, 1]$ as the power amplifier efficiency at ${\rm {AP}_\mathtt {I}}$, hence the total circuit power consumed at ${\rm {AP}_\mathtt {I}}$ is denoted by $P_{C}=\eta + P_{\rm AP_{\mathtt I}}+\sum_{k=1}^{K_{\rm E+I}} P_{{\rm D}_{k}}$,
where $P_{\rm AP_{\mathtt {I}}}$ denotes the overall static power consumption at the ${\rm {AP}_\mathtt {I}}$.
At the $k$th IoT device terminal, $P_{{\rm D}_{k}}$ indicates the hardware power dissipation. By assuming perfect CSI at all nodes, we perform EE maximization of ${\mathtt {I}^{net}}$ by optimization of the PS coefficient; $\varphi_k \ \{\forall k \in K_{\rm E+I}\}$, transmit BF vectors; ${{\textbf W}_{\mathtt I}}$ $\{{{\textbf W}_{\mathtt I}}=[{{\textbf w}_{{{1}}}},\dots,{{\textbf w}_{{{K_{\rm E+I}}}}}]\in\mathbb{C}^{M_{\mathtt {I}}\times K_{{\rm E+I}}} \}$, and employing optimized phase shifts of ${\mathtt {U}^{net}}$ with minimum transmit power constraint.
\par
Taking into consideration PS coefficients, individual data rate limitations of the devices, and the total transmit power budget, EE is obtained by dividing the sum rate of the devices by the total power consumption of the network. The EE optimization problem for ${\mathtt {I}^{net}}$ is represented as follows
    \begin{subequations}
        \begin{empheq}[left=]{align}
            &({\rm P_1}): \underset{\varphi,\textbf{W}_{\mathtt I},{\Theta}}{\max}\hspace{12pt}\frac {{\mathcal R_{k}^{{\mathtt {I}^{net}}}}}{1/\eta{{\sum_{k=1}^{K_{\rm E+I}}}}||{\textbf{w}}_{{\rm D}_{k}}||^{2}+P_C}\label{eq:14a}
            \\
            & \hspace{35pt} \text{s.t.} \hspace{14pt} \Gamma_{{\rm D}_{k}}^{\rm ID} \geq {\Gamma _{{\rm D}_{k,{\rm min}}}}, \ \forall k \in {K_{\rm E+I}},\label{eq:14b}\\ 
            &\hspace{62pt}{{\displaystyle\sum_{k=1}^{K_{\rm E+I}}}}||{\textbf{w}}_{{\rm D}_{k}}||^{2}\leq P_{{{\rm AP_{\mathtt {I}}}}}, \label{eq:14c}\\
            &\hspace{62pt}f(\mathscr E) \geq \frac{\bar E_{\mathcal H_k}}{{\varphi_k}\mu_k},\ \ \forall k \in K_{\rm E+I}, \label{eq:14d}
            \\
            &\hspace{62pt}|\Theta_{k,n,n}|=1,\ \ \forall n \in N,\label{eq:14e} 
            \\
         &\hspace{63pt}0\leq{\theta_{k,n}}\leq 2\pi, \ \ \forall n \in N,\label{eq:14f}
            \\
            &\hspace{64pt}0<{\varphi_{k}} < 1, \ \ \forall k \in K_{\rm E+I},\label{eq:14g}
        \end{empheq}
    \end{subequations}
where optimization variables are implicitly considered in objective function for notational brevity, and $f(\mathscr E)=\sum_{k=1}^{K_{\rm E+I}}{|(\textbf{h}_{{{\rm D}_{{\rm r},k}}}^H \Theta \textbf{G}_{\rm D}+\textbf{g}_{{{\rm D}_{{\rm d},k}}}^H)\textbf{w}_{{\rm D}_{k}}|^2}$ is used for notational simplicity. The constraint~(\ref{eq:14b}) ensures the QoS requirements, with $\Gamma_{{\rm D}_{{k},{\rm min}}} = 2^{{\mathcal R}_{{\rm D}_{k,{\rm min}}}}- 1$ shows the minimum SINR requirement of $k$-th IoT device for ID, and ${\mathcal R_{{\rm D}_{k,{\rm min}}}}$ represents the minimum data rate required. The maximum transmission power $P_{\rm AP_{\mathtt {I}}}$ at ${\rm AP}_{\mathtt I}$ is constrained by~(\ref{eq:14c}), where $||{\textbf w}_{{\rm D}_{k}}||^2$ and $\theta_{{n}}$ illustrate the respective Euclidean-norm of ${\textbf{w}}_{{\rm D}_{k}}$ and the $n$-th element of $\Theta$. The constraint~(\ref{eq:14d}) guarantees the EH requirements of the devices employing ${\bar E_{\mathcal H_k}}$ as the minimum amount of energy harvested. The constraints~(\ref{eq:14e}) and~(\ref{eq:14f}), respectively, define the passive nature of the reflecting elements of the IRS and the phase shifts range, while the constraint~(\ref{eq:14g}) represents the bound range of PS coefficients.
\par
Because of the non-convex objective function and coupled optimization variables, it appears clearly that $(\rm P_1)$ cannot be solved directly. Furthermore,~(\ref{eq:14d}) and~(\ref{eq:14f}) have PS relationship constraints that further complicate $(\rm P_1)$; because the feasible region of the PS ratio is not always non-empty. Therefore, to solve $(\rm P_1)$, an alternating optimization solution is proposed for the given framework by dividing the original optimization problem into sub-problems and solving them by employing effective phase cooperation between networks. To exploit the optimal phase shifts, we solve the problem by first optimizing passive phase shifts of $\mathtt U^{net}$ using AO and then using those optimized fixed phase shifts $\Theta$ to maximize EE at $\mathtt I^{net}$ using EH coefficients and active BF optimization.

\section{Alternating Optimization Solution for ${\mathtt {U}^{net}}$}
\label{sec:III}
To solve $(\rm P_1)$, an optimization problem is first formulated for ${\mathtt {U}^{net}}$ to obtain the optimal phase shifts. Hence, EE maximization problem similar to $(\rm P_1)$ is developed for the ${\mathtt {U}^{net}}$ and a solution is provided via an AO. In particular, we dissociate problem $(\rm P_1)$ into passive phase shifts optimization and active BF optimization sub-problems for ${\mathtt {U}^{net}}$ and address them alternatively. 
To solve $(\rm P_1)$ via AO, we employed low computational heuristic sub-optimal solutions, i.e., MMSE and ZFBF, to obtain the beamformers; ${{\textbf W}_{\mathtt U}}$ $\{{{\textbf W}_{\mathtt U}}=[{{\textbf w}_{{{1}}}},\dots,{{\textbf w}_{{{K_{\rm I}}}}}]\in\mathbb{C}^{M_{\mathtt {U}}\times K_{{\rm I}}} \}$ with closed-form solutions, while the passive phase shifts of ${\mathtt {U}^{net}}$ are optimized using a semi-definite programming (SDP) approach. Further, it is noticeable that IRS is deployed for ${\mathtt {U}^{net}}$; therefore, in the case of IRS-enabled ${\mathtt {U}^{net}}$, the total power consumed is given by $P_{C}=\eta + P_{\rm AP_{\mathtt U}} + P_{ N}+\sum_{k=1}^{K_{\rm I}} P_{{\rm R}_{k}}$, where the factor $P_{N}$ denotes the power consumption at the IRS terminal and $P_{{\rm R}_{k}}$ shows power dissipation at the $k$-th user device.
Hence, similar to $(\rm P_1)$ but only with the WIT technique, the optimization problem for ${\mathtt {U}^{net}}$ with only ID receivers is formulated as
          \begin{subequations}
        \begin{empheq}[left=]{align}
            &({\rm P_2}): \underset{\textbf{W}_{\mathtt U},{\Theta}}{\max}\hspace{12pt}\frac {{\mathcal R_{k}^{{\mathtt {U}^{net}}}}}{1/\eta{{\sum_{k=1}^{K_{\rm I}}}}||{\textbf{w}}_{{\rm R}_{k}}||^{2}+P_C}\label{eq:15a}
            \\
            & \hspace{30pt} \text{s.t.} \hspace{17pt} \Gamma_{{\rm R}_{k}} \geq {\Gamma _{{\rm R}_{k,{\rm min}}}}, \ \forall k \in {K_{\rm I}},\label{eq:15b}\\ 
            &\hspace{60pt}{{\displaystyle\sum_{k=1}^{K_{\rm I}}}}||{\textbf{w}}_{{\rm R}_{k}}||^{2}\leq P_{{{\rm AP_{\mathtt {U}}}}}, \label{eq:15c}\\
            &\hspace{62pt}|\Theta_{k,n,n}|=1,\ \ \forall n \in N,\label{eq:15d} 
            \\
        &\hspace{62pt}0\leq{\theta_{k,n}}\leq 2\pi, \ \ \forall n \in N,\label{eq:15e}
        \end{empheq}
    \end{subequations}
here, $\Gamma_{{\rm R}_{{k},{\rm min}}} = 2^{{\mathcal R_{{\rm R}_{k,{\rm min}}}}}- 1$ presents $k$-th IDR minimum SINR requirement, and ${\mathcal R_{{\rm R}_{k,{\rm min}}}}$ shows the minimum data rate required by ${\rm R}_{{k}}$ receiver. The factors $||{\textbf w}_{{\rm R}_{k}}||^2$ represent the Euclidean-norm of ${\textbf{w}}_{{\rm R}_{k}}$ and $\theta_{{n}}$ signify the $n$-th element of $\Theta$. The maximum transmission power available at ${\rm AP}_\mathtt{U}$ is denoted by $P_{\rm AP_{\mathtt{U}}}$. $(\rm P_2)$ is also computationally intractable owing to non-convex objective function and optimization variables provided in~(\ref{eq:15b}),~(\ref{eq:15c}), and~(\ref{eq:15e}) constraints. 
So, in order to solve $(\rm P_2)$, we perform AO by first exploiting linear BF optimization for the given phase shifts, and then phase optimization is performed for the obtained sub-optimal beamformers.
\subsection{Transmit Beamforming Optimization}
 \label{sec:III-A}
$(\rm P_2)$ is transformed into an EE maximization problem under the individual users data rate constraint, for a given $\Theta$. 
Hence, for ${\mathtt {U}^{net}}$ the $({\rm P_{2}})$ is reformulated into sub-problem as follows
  \begin{subequations}
        \begin{empheq}[left=]{align}
            &({\rm P_{2.1}}): \underset{\textbf{W}_{\mathtt U}}{\max}\hspace{12pt}\frac {{\mathcal R_{k}^{{\mathtt {U}^{net}}}}}{1/\eta{{\sum_{k=1}^{K_{\rm I}}}}||{\textbf{w}}_{{\rm R}_{k}}||^{2}+P_C}
            \label{eq:16a}\\
            & \hspace{35pt} \text{s.t.} \hspace{17pt}~(\ref{eq:15b}), ~(\ref{eq:15c}) 
        \end{empheq}
    \end{subequations}
The stated constraints allow $(\rm P_{2.1})$ to be tractable; as a result, the heuristic closed-form BF schemes are investigated below to obtain sub-optimal solution\footnote{Heuristic BF schemes discussed in Section~\ref{sec:III} are applied to both
${\mathtt {U}}^{net}$ and ${\mathtt {I}}^{net}$. Here, general notations are used, i.e., $k \in K_{\rm \{I,E+I\}}$.
}.
\par
${Heuristic \ Approach:}$
The SE and EE of B5G wireless communication networks are significantly enhanced by adaptive transmit BF~\cite{bengtsson2018optimum}. In literature, a multitude of exhaustive and heuristic BF schemes are investigated to provide optimal and sub-optimal solutions for different network architectures~\cite{8371237}. Therefore, to avoid the computational complexity of optimal transmit BF schemes, in this article we employ two heuristic BF approaches, i.e., transmit MMSE/regularized ZFBF and ZFBF, as our proposed framework is designed to facilitate low power SWIPT IoT applications and provide solution for the NP-hard multi-user transmit BF problem. Albeit these schemes provide sub-optimal solution, they are least complex, provide closed-form solutions, and can be easily tweaked to provide nearest optimal solution under special cases.
\par
Although $(\rm P_{2.1})$ is tractable, if the users required data rate increases or the ${\rm AP_{\mathtt {U}}}$ transmit power lowers, the maximization problem becomes infeasible for a given $\Theta$ under the constraint~(\ref{eq:15b}). 
To solve this, $(\rm P_{2.1})$ is reformulated as a power minimization problem given as
\begin{subequations}
        \begin{empheq}[left=]{align}
            &({\rm P_{2.1}^{*}}): \underset{\textbf{W}_{\mathtt U}}{\min}
            ~~\hspace{5pt}{{\displaystyle\sum_{k=1}^{K_{\rm I}}}}||{\textbf{w}}_{{\rm R}_{k}}||^{2}
              \label{eq:17a}\\
             & \hspace{35pt} \text{s.t.} \hspace{14pt} \Gamma_{{\rm R}_{k}} \geq {\Gamma _{{\rm R}_{k,{\rm min}}}}, \ \forall k \in {K_{\rm I}}.\label{eq:17b}
        \end{empheq}
            \end{subequations}
$({\rm P_{2.1}^{*}})$ clearly contains a convex objective function and can be traced using different convex approximation techniques including second order cone program (SOCP)~\cite{8959161,9741332}, semidefinite programming (SDP)~\cite{luo2010semidefinite,anjos2011handbook} under the minimum transmit power, or a fixed point iteration algorithm~\cite{schubert2004solution,huang2009rank}. However, to solve the hidden convexity of constraint~(\ref{eq:15b}) we employed inner product property of phase rotation from~\cite{bengtsson2018optimum}, which states that for absolute values of SINRs, the inner product of signal power is positive and real valued, i.e., $\sqrt{{|(\textbf{h}_{{{\rm R}_{{\rm r},k}}}^H \Theta \textbf{G}_{\rm R}+\textbf{g}_{{{\rm R}_{{\rm d},k}}}^H)\textbf{w}_{{\rm R}_{k}}|^2}}={(\textbf{h}_{{{\rm R}_{{\rm r},k}}}^H \Theta \textbf{G}_{\rm R}+\textbf{g}_{{{\rm R}_{{\rm d},k}}}^H)\textbf{w}_{{\rm R}_{k}}}\geq 0$. Additionally, let's define two-fold channel gains as follows for ease of representation: $\textbf{f}_{{{\rm R}_{k}}}^H={(\textbf{h}_{{{\rm R}_{{\rm r},k}}}^H \Theta \textbf{G}_{\rm R}+\textbf{g}_{{{\rm R}_{{\rm d},k}}}^H)}$. Hence, using inner product property the constraint~(\ref{eq:15b}) can be reformulated as~\cite{bjornson2013optimal}
\begin{align}
{\frac{1}{{\Gamma _{{\rm R}_{k,{\rm min}}}}{\sigma_{k}^2}}}|\textbf{f}_{{{\rm R}_{k}}}^H\textbf{w}_{{\rm R}_{k}}|^2\geq{\displaystyle\sum_{i\neq k}}{\frac{1}{\sigma_{k}^2}}|\textbf{f}_{{{\rm R}_{k}}}^H\textbf{w}_{{\rm R}_{k}}|^2+1  
\Leftrightarrow \notag \\ 
{\frac{1}{\sqrt{{{\Gamma _{{\rm R}_{k,{\rm min}}}}{\sigma_{k}^2}}}}}{\mathds R}(\textbf{f}_{{{\rm R}_{k}}}^H\textbf{w}_{{\rm R}_{k}})\geq\sqrt{{\displaystyle\sum_{i\neq k}}{\frac{1}{\sigma_{k}^2}}|\textbf{f}_{{{\rm R}_{k}}}^H\textbf{w}_{{\rm R}_{k}}|^2+1},
\label{eq:18}
\end{align}
here, ${\mathds R}\{\cdot\}$ shows the real part; so,~(\ref{eq:18}) provides a reformulated SINR that is a convex second order cone constraint~\cite{dahrouj2010coordinated,bengtsson2018optimum}. Now, by using this new constraint~(\ref{eq:18}), $({\rm P_{2.1}^{*}})$ can be solved feasibly using convex optimization theory~\cite{bengtsson2018optimum,bjornson2013optimal}. Specifically, the strong duality and Karush Kuhn Tucker (KKT) conditions are sufficient to solve $({\rm P_{2.1}^{*}})$~\cite{yu2007transmitter}. Hence, by employing strong duality and KKT conditions the Lagrangian function defined for $({\rm P_{2.1}^{*}})$ is obtained as
\begin{align}
    \mathcal{L}({{\textbf W}_{\mathtt I}}, {{\Lambda}_{\mathtt I}})={{\displaystyle\sum_{k=1}^{K_{\rm I}}}}||{\textbf{w}}_{{\rm R}_{k}}||^{2}+{{\displaystyle\sum_{k=1}^{K_{\rm I}}}}{{\lambda}_{k}}\Big({\displaystyle\sum_{i\neq k}}{\frac{1}{\sigma_{k}^2}}|\textbf{f}_{{{\rm R}_{k}}}^H\textbf{w}_{{\rm R}_{i}}|^2+1\notag \\
    -{\frac{1}{{\Gamma _{{\rm R}_{k,{\rm min}}}}{\sigma_{k}^2}}}|\textbf{f}_{{{\rm R}_{k}}}^H\textbf{w}_{{\rm R}_{k}}|^2\Big),
    \label{eq:19}
\end{align}
where ${{\Lambda}_{\mathtt I}}=[{{\lambda}_{{{1}}}},\dots,{{\lambda}_{{{K_{\rm I}}}}}]$, $\lambda_{k}\geq0$ $(k\in\{1,\dots, K_{\rm I}\})$ is the Lagrangian multiplier associated with the $k$-th SINR constraint and can be computed using convex optimization or fixed point equations~\cite{gershman2010convex}. The dual function is ${\rm min}_{{\textbf w}_{{1}},{\textbf w}_{{2}},\ldots,{\textbf w}_{{K_{\rm I}}}}\mathcal{L}=\sum_{k=1}^{K_{\rm I}}{{{{{\lambda_{{k}}}}}}}$ and the strong duality property of [$(\rm P_{2.2})$,~\cite{bjornson2014optimal}] implies that the total power is $\sum_{k=1}^{K_{\rm I}}||{\textbf{w}}_{{{\rm R}_k}}||^2$. 
By exploiting KKT conditions from~\cite{yu2007transmitter} and using a power allocation scheme discussed in~\cite{bjornson2013optimal} (Theorem 3.16), the optimal BF vector for $({\rm P_{2.1}^{*}})$ under given $\Theta$ is expressed as
\begin{equation}
    {\textbf {w}}_{{\rm R}_k}^{*}={\sqrt{{P}_{k}}}\bar{\textbf{w}}_{{\rm R}_k}^{*},\ \ {\rm where} \ \
\bar{\textbf{w}}_{{\rm R}_k}^{*}=\frac{{\Bigl({\textbf {I}}_{M_\mathtt {U}}+\displaystyle\sum_{i=1}^{K_{\rm I}} {\frac {\lambda_{i}}{\sigma_{k}^{2}}}}{\textbf{f}_{{{\rm R}_{i}}}} {\textbf{f}_{{{\rm R}_{i}}}^H}\Bigl)^{-1}{\textbf{f}_{{{\rm R}_{k}}}}}{\Bigl\lVert {{\Bigl({\textbf {I}}_{{M_\mathtt {U}}}+\displaystyle\sum_{i=1}^{K_{\rm I}} {\frac {\lambda_{i}}{\sigma_{k}^{2}}}}{\textbf{f}_{{{\rm R}_{i}}}} {\textbf{f}_{{{\rm R}_{i}}}^H}\Bigl)^{-1}{\textbf{f}_{{{\rm R}_{k}}}}}\Bigl\rVert },
\label{eq:20} 
\end{equation}
where, the factor $P_k$ shows the BF power at the ${\rm {AP}_\mathtt {U}}$ and $\bar{\textbf{w}}_{{\rm R}_k}^{*}$ signify the uniformed BF direction for $k$-th user. Since we obtain the BF directions, the unknown power factors associated with $K_{\rm I}$ users can be obtained by using the equality condition of constraint~(\ref{eq:18}) at the optimal solution. Hence, for $K_{\rm I}$ known BF linear equations, the corresponding power factors can be obtained as~\cite{bjornson2014optimal}
\begin{align}
\begin{bmatrix} 
{P}_{1} \\
\vdots \\
{P}_{K_{\rm I}}  \\
\end{bmatrix}
=\mathbf A^{-1} \begin{bmatrix} 
{\sigma}_{1} \\
\vdots \\
{\sigma}_{K_{\rm I}}  \\
\end{bmatrix} \ {\rm where} \
[\mathbf A]_{ij}&= 
\begin{cases}
{{\frac{1}{{\Gamma _{{\rm R}_{k,{\rm min}}}}}}|\textbf{f}_{{{\rm R}_{k}}}^H \bar {\textbf{w}}_{{\rm R}_{k}}|^2}, \ i = j, \\
-{|\textbf{f}_{{{\rm R}_{k}}}^H \bar{\textbf{w}}_{{\rm R}_{k}}|^2}, \hspace{0.8cm} i \neq j, \\
\end{cases}. 
\label{eq:21}
\end{align}
Now,~(\ref{eq:20}) and~(\ref{eq:21}) combined provide an optimal closed-form BF solution for $({\rm P_{2.1}^{*}})$ as a function of simple Lagrange multipliers. Using this optimal BF structure obtained by solving $({\rm P_{2.1}^{*}})$, we can easily solve our original EE optimization problem, i.e., $({\rm P_{2.1}})$. The solution to $({\rm P_{2.1}^{*}})$, provides optimal BF vectors under minimum transmit power; therefore, this solution can feasibly satisfy the constraint~(\ref{eq:15b}) of $({\rm P_{2.1}})$. Further, using optimal BF vectors of $({\rm P_{2.1}^{*}})$, the optimal BF solution for $({\rm P_{2.1}})$ is provided as
\begin{equation}
    {\textbf {w}}_{{\rm R}_k}^{*}={\sqrt{{P}_{k}}}\frac{{\Bigl({\textbf {I}}_{M_\mathtt{U}}+\displaystyle\sum_{i=1}^{K_{\rm I}} {\frac {\lambda_{i}}{\sigma_{k}^{2}}}}{\textbf{f}_{{{\rm R}_{i}}}} {\textbf{f}_{{{\rm R}_{i}}}^H}\Bigl)^{-1}{\textbf{f}_{{{\rm R}_{k}}}}}{\Bigl\lVert {{\Bigl({\textbf {I}}_{{M_\mathtt{U}}}+\displaystyle\sum_{i=1}^{K_{\rm I}} {\frac {\lambda_{i}}{\sigma_{k}^{2}}}}{\textbf{f}_{{{\rm R}_{i}}}} {\textbf{f}_{{{\rm R}_{i}}}^H}\Bigl)^{-1}{\textbf{f}_{{{\rm R}_{k}}}}}\Bigl\rVert },\ \forall k \in {K_{\rm I}}.
\label{eq:22}
\end{equation}
\par
For $K_{\rm I}$ users the matrix inversion in~(\ref{eq:22}) remains same, hence the optimal BF matrix for ${\mathtt {I}^{net}}$ can be represented as ${{\textbf W}_{\mathtt U}^{*}}$ $\{{{\textbf W}_{\mathtt U}^{*}}=[{{\textbf w}_{{{1}}}^{*}},\dots,{{\textbf w}_{{{K_{\rm I}}}}^{*}}]\in\mathbb{C}^{M_{\mathtt {U}}\times K_{{\rm I}}} \}$ and the channel matrix for IRS-enabled network as ${\textbf F}_{\mathtt {U}} = [{\textbf f}_{1},{\textbf f}_{2},\ldots,{\textbf f}_{K}]\in\mathbb{C}^{M_{\mathtt {U}}\times K_{{\rm I}}}$, where $\textbf{f}_{{{\rm R}_{k}}}^H={(\textbf{h}_{{{\rm R}_{{\rm r},k}}}^H \Theta \textbf{G}_{\rm R}+\textbf{g}_{{{\rm R}_{{\rm d},k}}}^H)}$. The resultant combined channel response can be expressed as $\sum_{i=1}^{K_{\rm I}} {\frac {\lambda_{i}}{\sigma_{k}^{2}}}{\textbf{f}_{{{\rm R}_{i}}}} {\textbf{f}_{{{\rm R}_{i}}}^H}={\frac{1}{\sigma^2}}{\textbf{F}_{\mathtt {U}}}\Lambda{\textbf{F}^{H}_{\mathtt {U}}}$. Thus, it is feasible to derive the compact BF matrix as 
  \begin{align}
{{\textbf W}_{\mathtt U}^{*}}&= \left({\textbf {I}}_{M_\mathtt {U}}+{\frac{1}{\sigma^2}}{\textbf{F}_{\mathtt {U}}}\Lambda{\textbf{F}^{H}_{\mathtt {U}}}\right)^{-1}{\textbf{F}_{\mathtt {U}}}{{\textbf P}^{\frac{1}{2}}},
\label{eq:23}
\end{align}
where ${{\Lambda}}={\rm diag}({{\lambda}_{{{1}}}},\dots,{{\lambda}_{{{K_{\rm I}}}}})$, is a diagonal matrix having ${K_{\rm I}}$ $\lambda$-parameters. ${{\textbf P}^{\frac{1}{2}}}$ shows the square root of power allocation matrix given as ${{\textbf{P}}}={\rm diag}{({ P}_{1}}/||({\textbf {I}}_{M_\mathtt {U}}+{\frac{1}{\sigma^2}}{\textbf{F}_{\mathtt {U}}}\Lambda{\textbf{F}^{H}_{\mathtt {U}})^{-1}{\textbf{f}_{1}}}
||^2,\cdots,{{ P}_{K_{\rm I}}}/||({\textbf {I}}_{M_\mathtt {U}}+{\frac{1}{\sigma^2}}{\textbf{F}_{\mathtt {U}}}\Lambda{\textbf{F}^{H}_{\mathtt {U}}})^{-1}{\textbf{f}_{K_{\rm I}}||^2})$.
So, the optimum BF from~(\ref{eq:23}) will provide the optimal SINR, which in result maximizes $\rm(P_{2.1})$. By employing~(\ref{eq:23}) the closed-form BF solutions for ZFBF and transmit MMSE/Regularized ZFBF schemes are described below. 

\subsubsection{ZFBF}
ZFBF is a potential technique for reducing inter-user interference~\cite{wiesel2008zero,1468466}. It is referred to as the channel inversion scheme since it contains the channel matrix's pseudo-inverse ${\textbf{F}^{H}_{\mathtt {U}}}$, i.e., ${\textbf{F}_{\mathtt {U}}}({\textbf{F}^{H}_{\mathtt {U}}}{\textbf{F}_{\mathtt {U}}})^{-1}$. When using ZFBF, interference power is reduced while SINR is optimized. By cancelling noise and interference, this criteria decouples the selection of the BF direction. Hence, using asymptotic properties from~\cite{bjornson2013optimal}, the relation for ZFBF is obtained as 
  \begin{align}
  {\textbf W}_{{\mathtt {U}}}^{*(\rm ZFBF)}&= {\textbf{F}_{\mathtt {U}}}({\textbf{F}^{H}_{\mathtt {U}}}{\textbf{F}_{\mathtt {U}}})^{-1}{\Lambda^{-1}}{ \bar{\textbf P}},
  \label{eq:24} 
  \end{align}
where ${\bar{{\textbf{P}}}}$ shows the modified power allocation matrix provided as
${\bar{\textbf{P}}}={\rm diag}{({ P}_{1}}/||({\sigma^2}{\textbf {I}}_{K_{\rm I}}+\Lambda{\textbf{F}^{H}_{\mathtt {U}}{\textbf{F}_{\mathtt {U}}})^{-1}{\textbf{f}_{1}}}
||^2,\cdots,{{ P}_{K_{\rm I}}}/||({\sigma^2}{\textbf {I}}_{K_{\rm I}}+\Lambda{\textbf{F}^{H}_{\mathtt {U}}}{\textbf{F}_{\mathtt {U}}})^{-1}{\textbf{f}_{K_{\rm I}}||^2})$. At large SNRs, ZFBF is considered to be the asymptotic optimal option; nevertheless, performance degrades at moderate SNRs.
\subsubsection{Transmit MMSE/Regularized ZFBF}
 \label{sec:III-B}
In a multi-user scenario, MMSE is acclaimed as the sub-optimal heuristic solution because it can overcome ZFBF's asymptotic optimality and provide better results at intermediate SINRs. Hence, using~(\ref{eq:20}) and~(\ref{eq:21}) and setting ${{\lambda}_k=\lambda, \ \forall k \in {{{K_{\rm I}}}}}$ the subsequent expression for MMSE optimal BF is obtained as
  \begin{align}
{\textbf W}_{\mathtt {U}}^{*(\rm MMSE)}&= {\textbf{F}_{\mathtt {U}}}\left({\textbf{I}_{K_{\rm I}}}+{\frac{\lambda}{\sigma^2}}{{\textbf{F}_{\mathtt {U}}^{H}}}{\textbf{F}_{\mathtt {U}}}\right)^{-1}{{\textbf P}^{\frac{1}{2}}}.
\label{eq:25}
\end{align}
Expression~(\ref{eq:25}) is also named as regularized ZFBF~\cite{joham2005linear,biglieri2007mimo} as it contain an identity matrix which acts as a regularization matrix in~(\ref{eq:25}). Here, $\lambda$ is referred as the regularization parameter that satisfies the sum property of transmission power, i.e., $\sum_{i=1}^{K_{\rm I}} {{\lambda_{i}}}=P$ with $\lambda=P/K$. Consequently, using a standard line search method,~(\ref{eq:25}) provides an optimal solution for a particular transmission case by adjusting the $\lambda$-parameter.
\subsection{Passive Phase Optimization}
\label{sec:III-B}
By utilizing the optimal transmit beamformers $\textbf W_{\mathtt U}^*$ obtained in Section~\ref{sec:III-A}, the phase shifts optimization can be solved for $({\rm P_2})$. The reformulated sub-problem is given as
          \begin{subequations}
        \begin{empheq}[left=]{align}
            &({\rm P_{2.2}}): \underset{{\Theta}}{\max}\hspace{12pt}\frac {{\mathcal R_{k}^{{\mathtt {U}^{net}}}}}{1/\eta{{\sum_{k=1}^{K_{\rm I}}}}||{\textbf{w}}_{{\rm R}_{k}}||^{2}+P_C}\label{eq:26a}
            \\
            & \hspace{35pt} \text{s.t.} \hspace{17pt}~(\ref{eq:15b}),~(\ref{eq:15d}),~(\ref{eq:15e})
            \label{eq:26b} 
        \end{empheq}
    \end{subequations}
The non-convex constraints make $(\rm P_{2.2})$ NP-hard optimization problem; however, by making a few variable changes, the problem can be feasibly transformed into a tractable form.
Let $\textbf{h}_{{{\rm R}_{{\rm r},k}}}^H \Theta \textbf{G}_{\rm R}\textbf{w}_{{\rm R}_{k}}^{*}={\bm \nu}^H {\textbf a}_{i,k}$ and $\textbf{h}_{{{{\rm D}_{{\rm d},k}}}}^H{\textbf w}_{{\rm R}_k}^*={b_{i,k}}$ where ${{\bm \nu}}=[\nu_1,\dots,\nu_{N}]^H$ for $\nu_n=e^{j\theta_n}, \forall n\in N$ and ${{\textbf a}_{i,k}}=(\textbf{h}_{{{{\rm R}_{{\rm r},k}}}}^H)\Theta \textbf{G}_{\rm R} {\textbf w}_{{\rm R}_{k}}^{*} \in \mathbb C^{N_{\rm IRS} \times 1}$.
Also, it is notable that although the same IRS phase shifts matrix, $\Theta$, is shared by all the users, the SINR constraints in~(\ref{eq:15b}) are not always mandated to be satisfied for a feasible solution of $(\rm P_ {2.2})$. As a result, a slack variable, $\beta_k$, is introduced to deal with the ${K_ {\rm I}}$ residual SINR. The reformulated sub-problem is stated as following by substituting new variables
   \begin{subequations}
        \begin{empheq}[left=]{align}
            &({\rm P_{2.2}^*}): \hspace{10pt}\underset{{\bm \nu},{\{{\beta}_{k}\}_{\scaleto{k=1}{4pt}}^{\scaleto{K_{\rm I}}{4pt}}}}{\rm max}
         ~~\hspace{7pt}{{{\displaystyle\sum_{k\in K_{\rm I}}}}}\beta_{k}
              \label{eq:27a}\\
            &\hspace{5pt} \text{s.t.} \hspace{5pt} 
            \frac{|{\bm \nu}^H {\textbf a}_{k,k}+{b}_{k,k}|^2}{\sum_{i\neq k}^{K_{\rm I}}{|{\bm \nu}^H {\textbf a}_{i,k}+{b}_{i,k}|^2+\sigma_{{k}}^2}}\geq \Gamma_{{\rm R}_{{k,\rm min}}}+{\beta_{k}},\ \forall k,
              \label{eq:27b}   \\   
               &\hspace{20pt} |\nu_{n}|=1, \ \ \forall n \in N,\ \ \beta_{k} \geq 0, \ \forall k \in {K_{\rm I}}.
                \label{eq:27c}
        \end{empheq}
        \label{eq:16}
    \end{subequations}
where~(\ref{eq:15d}) transforms into a non-convex unit-modulus constraint, i.e., ${|{ \nu_{n}}|^2=1}, \forall n$. Although $(\rm P_{2.2^*})$ being a non-convex optimization problem, the fully separable phase shifts allow the constraints to be expressed in quadratic form given as
\begin{equation}
\textbf X_{i,k}=
 \begin{bmatrix}
  {\textbf a_{i,k}}{\textbf a_{i,k}^{H}} & {\textbf a_{i,k}}{ b_{i,k}^{H}}\\ 
  {\textbf a_{i,k}^{H}}{ b_{i,k}} & 0
\end{bmatrix},
\hspace{10pt}\bar {\bm \nu}=
 \begin{bmatrix}
 {\bm \nu} \\ 
  1 
\end{bmatrix}. \notag
\end{equation}
\par
By using algebraic manipulations, we obtain $\bar {\bm \nu}^{H}\textbf X_{i,k}\bar {\bm \nu}={\rm \textbf Tr}(\textbf {VX})$ where $\textbf {V}=\bar{\bm \nu}\bar {\bm \nu}^{H}$. The conditions $\textbf {V}\succeq 0$ and rank$(\textbf {V})=1$ should be satisfied by $\textbf {V}$. Since the rank-one constraint is non-convex and makes the problem $(\rm P_{2.2}^{*})$ arduous to tackle, we drop this constraint by employing the SDR approach and relax the resultant problem as
\par
\setlength{\abovedisplayskip}{1pt}
\setlength{\belowdisplayskip}{1pt}
        \begin{subequations}
        \begin{empheq}[left=]{align}
            &{(\rm P_{2.2}^{**})}: \underset{{\textbf V\in \mathbb{H}^{n}},{\{{\beta}_{k}\}_{\scaleto{k=1}{4pt}}^{\scaleto{K_{\rm I}}{4pt}}}}{\rm max}
         ~~\hspace{7pt}{\displaystyle\sum_{k\in K_{{\rm I}}}}\beta_{k}
             \label{eq:28a}\\
            & \hspace{2pt} \text{s.t.}
            \hspace{8pt}
            {\rm \textbf Tr}(\textbf {V}\textbf X_{k,k})
            +|{b_{k,k}}|^{2} \geq \Gamma_{{\rm R}_{k,\rm min}}\Bigg[\Bigg\{\sum_{i\neq k}^{K_{\rm I}}{\rm \Big(\textbf Tr}(\textbf{V}\textbf X_{i,k}) \notag \\ 
            &\hspace{33pt} +|{b_{i,k}}|^{2}\Big)\Bigg\}
            +\sigma_{{k}}^2\Bigg]+{\beta_{k}}, \ \forall k \in K_{\rm I},
             \label{eq:28b}\\ 
           & \hspace{25pt} {\rm {diag}}(\textbf {V})=[{\textbf I]_{{N+1\times 1}}},
           \label{eq:28c} \\
           & \hspace{25pt}
            \textbf {V}\succeq 0,\ \ \  {\beta_{k}}\geq 0, \  \ \forall k \in K_{\rm I},
            \label{eq:28d}
        \end{empheq}
    \end{subequations}
where ${{\textbf V\in \mathbb{H}^{n}}}$ guarantees $\textbf {V}\succeq 0$. ${(\rm P_{2.2}^{**})}$ is now a standard semi-definite convex programming (SDP) problem that can be solved optimally using the Matlab-based modelling tool CVX for convex optimization~\cite{grant2014cvx}. However, ${(\rm P_{2.2}^{**})}$ does not always provide a rank-one solution, so a Gaussian randomization approach can be used to obtain this solution~\cite{huang2009rank,5447068}. In particular, we first perform the eigenvalue decomposition of $\textbf {V}$ as $\textbf {V}=\textbf {U}\sum\textbf {U}^{H}$.
Afterwards, we obtain the modified solution to $({\rm P_{2.2}^*})$ as $\bar {\bm \nu}=\textbf {U}\sum^{\frac{1}{2}}\textbf {r}$, where $\textbf {r}\sim \mathcal{CN}(0,\textbf {I}_{N+1}) \in \mathbb{C}^{(N+1) \times 1}$ is a random vector selected from a large number of random generalized circularly symmetric complex Gaussian (CSCG) vectors for maximizing the objective function of $({\rm P_{2.2}^*})$. The resultant solution of the problem $({\rm P_{2.2}^*})$ is obtained
by ${\bm \nu}^{*} ={e^{j{\rm arg}}}({\bar{\bm \nu}}_{[1:N]}/{\bar {\bm \nu}}_{[N+1]})$ which satisfies the constraint~(\ref{eq:27c}). The resulting approach offers at least $\pi/4$-approximation of the optimal values for a particular objective function over a large number of randomizations.
\subsection{Computational Complexity Analysis of AO}
 \label{sec:III-C}
Algorithm~\ref{alg:alg1} summarizes the AO algorithm for addressing the problem $(\rm P_2)$. The AO method described in Section~\ref{sec:III-A} and~\ref{sec:III-B} is an iterative, multi-stage optimization algorithm. To optimize transmit beamformers and phase shifts, there are two sub-problems in the outer loop, and each sub-problem needs to be solved using an iterative update approach, respectively. In particular, the MMSE algorithm requires inverse matrix operations with complexity $\mathcal{O}\big(K_{\rm I}M^{3}\big)$. Additionally, a one-dimensional search (using standard line search) for $\lambda$ is also required. Therefore, MMSE algorithm complexity is illustrated as $\mathcal{O}\big(I_{\lambda}(K_{\rm I}M^{3}\big)\big)$, where $I_{\lambda}$ show the iteration numbers required to find $\lambda$. For phase shifts optimization, the problem $({\rm P_ {2.2}})$ optimizes the IRS phase shifts at each iteration by relaxing SDP problem using an interior point approach. Thus, the computational complexity of the problem $({\rm P_{2.2}})$ in solving the SDP problem can be expressed as $\mathcal{O}\big(({N+1})^{3.5}\big)$~\cite{9133435}. The total complexity of solving Algorithm~\ref{alg:alg1} is $\mathcal{O}\big(I_{out}\big(I_{\lambda}(K_{\rm I}M^{3}\big)+I_{inn}({N+1})^{3.5}\big)\big)$, where $I_{inn}$ and $I_{out}$ show inner and outer iterations to reach convergence, respectively. 
\begin{algorithm}[!t]
\caption{{ {AO Algorithm for Solving $(\rm P_2)$}}}
\label{alg:alg1}
\begin{algorithmic}
\STATE 
\small
\STATE {\textsc{\textbf{Initialization}:}} $r$: the iteration number, ${\xi>0}$: convergence accuracy, $\textbf w_{{\rm R}_{k}}^{(0)}$, ${\bm \nu}^{(0)}$: initial feasible solutions;

\STATE {\textsc{\textbf{Calculate}}} the initial objective value of problem $(\rm P_2)$;

\STATE {\textsc{\textbf{Perform:}}} {\textbf {AO algorithm}};

\STATE \hspace{0.3cm}1. Given $\bm \nu^{(0)}$ , $\textbf w_{{\rm R}_{k}}^{(r)}$: Calculate $\textbf w_{{\rm R}_{k}}^{(r+1)}$ by solving problem \STATE \hspace{0.8cm} $(\rm P_{2.1})$  with Heuristic Approach;

\STATE \hspace{0.3cm}2. Given $\textbf w_{k}^{(r+1)}$ and ${\bm \nu}^{(r)}$: 
 Calculate ${\bm \nu}^{(r+1)}$ by solving 
 \STATE \hspace{0.8cm}
 problem $(\rm P_{2.2}^{**})$ with SDR Approach;

 \STATE \hspace{0.7cm}i. Obtain \textbf{V} by solving $(\rm P_{2.2}^{**})$ using CVX;
 
 \STATE \hspace{0.6cm}ii. Perform Gaussian randomization to obtain $\bar {\bm \nu}=\textbf {U}\sum^{\frac{1}{2}}\textbf {r}$;

 \STATE \hspace{0.55cm}iii. Obtain phase shifts using ${\bm \nu} ={e^{j{\rm arg}}}({\bar{\bm \nu}}_{[1:N]}/{\bar {\bm \nu}}_{[N+1]})$

\STATE \hspace{0.3cm}3. \textbf {Update:} transmit beamformers ${\textbf W}_{\mathtt {U}}^{*}$ and phase shifts ${\bm \nu}^{*}$;

\STATE {\textsc{\textbf {Repeat}}} $r=r+1$ until $(\rm P_2)$ objective value falls below a threshold accuracy i.e.,
~(\ref{eq:15a})$\leq \xi$.
\end{algorithmic}
\end{algorithm}     
\section{Low Complexity Alternative Solution}
\label{sec:IV}
Although the AO approach presented in the above section provides a high-quality converging solution for ${(\rm P_{2.2})}$, it is computationally complex due to the SDR scheme, and this complexity becomes more significant with increasing IRS elements. This section presents an alternative, low-complexity solution to this problem by decoupling the active beamformers and IRS phase shifts design. The motivation behind this idea is to take advantage of IRS's short-range/local coverage, which allows users lying in the vicinity of an IRS-enabled network to receive reflected signals at their respective ends. Motivated by this, we can separately optimize IRS phase shifts and transmit beamformers for all users while satisfying their QoS requirements.
\subsection{Phase Shifts Optimization}
 \label{sec:IV-A}
By employing the sum of all users effective IRS channel gains and leveraging the variable changes made in Section~\ref{sec:III-B}, we reformulate the optimization problem ${(\rm P_{2.2})}$ and use an iterative approach to solve for phase optimization. The reformulated sub-problem by substituting variables is given by
   \begin{subequations}
        \begin{empheq}[left=]{align}
            &({\rm P_{2.3}}): \hspace{10pt}\underset{\bm \nu}{\rm max}
          ~~\hspace{7pt}{\displaystyle\sum_{k\in {K_{\rm I}}}{||({\bm \nu}^{H}\textbf a_{{\rm r},k}+{{\textbf d}_{{\rm d}, k}^H)}||^2}}
              \label{eq:29a}\\
            &\hspace{45pt} \text{s.t.} \hspace{20pt}|\nu_{n}|=1, \ \ \forall n \in N,
              \label{eq:29b}    
        \end{empheq}
    \end{subequations}
where ${{\bm \nu}}=[\nu_1,\dots,\nu_{N}]^H$ for $\nu_n=e^{j\theta_n}, \forall n\in N$, $\textbf a_{{\rm r},k}={\rm diag}(\textbf{h}_{{{\rm R}_{{\rm r},k}}}^H)\textbf{G}_{\rm R}\in \mathbb C^{1{\times M_{\mathtt U}}}$, and $\textbf{g}_{{{\rm R}_{{\rm d},k}}}^{H}={{\textbf d}_{{\rm d},k}^{H}}$.
The constraint~(\ref{eq:15e}) is now transformed into a unit-modulus constraint, i.e., ${|{ \nu_ {n}}|^2=1}, \forall n$. $(\rm P_{2.3})$ is still a non-convex optimization problem due to constraint~(\ref{eq:29b}). However, the phase optimization can be solved by an iterative method due to the fully separable phase shifts in constraint~(\ref{eq:29b}). Thus, we exploit low complexity EBCD approach for solving $(\rm P_ {2.3})$ by iteratively optimizing each element of ${\bm \nu}$, i.e., ${\nu}_{n}\ (\forall n \in N)$ given the other phase shifts, ${\nu}_{l}\ (\forall l \in N)$ where $l\neq n$ as fixed. The sub-problem is presented as
   \begin{subequations}
        \begin{empheq}[left=]{align}
            &({\rm P_{2.3}^{*}}): \underset{\bm \nu}{\rm max} \ \ f({\bm \nu)}
              \label{eq:30a}\\
            &\hspace{37pt} \text{s.t.} \hspace{10pt}| \nu_{n}|=1, \  \forall n \in N.
              \label{eq:30b}           
        \end{empheq}
    \end{subequations}
    \par
With fixed ${\nu}_{l}$, $(\rm P_ {2.3})$ becomes a linear objective function given as
\begin{align}
    f({\bm \nu)=}{{{{\sum_{l\neq n}^N}{\nu}_{l}\textbf{Q}{(l,l)} {\nu}_{l}^{H}}+2\mathbb{R}\{{\nu}_{n}\vartheta_n\}+S}}
              \label{eq:31}
\end{align}
where
\begin{align}
{\textbf{Q}}&={\sum_{k}{\textbf a_{{\rm r},k}}{\textbf a_{{\rm r},k}^{H}}},\ \forall {{k \in {K_{\rm I}}}},
\label{eq:32}\\
{\vartheta_{n}}&={\sum_{l\neq n}^{N}{\textbf{Q}{(n,l)}\nu_{l}^{H}}}-\tilde{\bm \vartheta}{(n)},
\label{eq:33} \\
\tilde{\bm \vartheta}&= {{\sum_{k}{{\textbf{a}}_{{\rm r},k}}}}{{\textbf d}_{{\rm d},k}^{H}}, \ \forall {k \in {K_{\rm I}}}.
\label{eq:34}\\
S&=\textbf{Q}{(n,n)}-2\mathbb{R}{\sum_{l\neq n}^{N}}\nu_{l}\tilde{\bm \vartheta}(l)+{\sum_{k}|{{\textbf d}_{{\rm d},k}^{H}}|^{2}}, \ \forall {k \in {K_{\rm I}}}.
\label{eq:35} 
\end{align}
By dropping constant term $S$ from~(\ref{eq:31}) and taking into account the unit-modulus constraint with some algebraic properties, the sub-problem $({\rm P_{2.3}^{*}})$ becomes the given update rule,
\setlength{\abovedisplayskip}{4pt}
\setlength{\belowdisplayskip}{4pt}
   \begin{subequations}
        \begin{empheq}[left=]{align}
            &({\rm P_{2.3}^{**}}): \underset{\bm \nu_{n}}{\rm max}
            ~~{{{2\mathbb{R}\{{\nu}_{n}\vartheta_n\}}}}
              \label{eq:36a}\\
            &\hspace{37pt} \text{s.t.} \hspace{10pt}| \nu_{n}|=1, \  \forall n \in N.
              \label{eq:36b}           
        \end{empheq}
    \end{subequations}
Consequently, by fixing $\nu_{l}$ $\{\forall l \in N\}, l\neq n$ the optimal solution to $({\rm P_{2.3}^{**})}$ becomes
\begin{align}
{\nu}_{n}^{*}&=
\begin{cases}
1, \hspace{2.6cm} {\rm if} \ {\nu}_{n}=0, \\
\frac{{-\vartheta}_{n}^{H}}{|{\vartheta_n|}}, \  \forall n \in N, \hspace{0.8cm} {\rm otherwise}.
\end{cases} 
\label{eq:37}
\end{align}
According to~(\ref{eq:37}), each block of $N-1$ phase shifts can be optimized iteratively by fixing the other $N$ phases until the convergence is achieved in this block, which is implied by the fractional decrease in $({\rm P_{2.3}^{*})}$ objective function value below a positive tolerance threshold. Algorithm~\ref{alg:alg2} summarizes the EBCD method for solving problem $(\rm P_{2.2})$ to obtain optimal phase shifts.
\begin{algorithm}[!t]
\small
\caption{EBCD Algorithm for Solving $(\rm P_{2.3})$}\label{alg:alg2}
\begin{algorithmic}
\STATE 
\STATE {\textbf{Input:}} ${\bm \nu}_{n}^{(0)}$: initial feasible solution, $r$: iteration number, ${\xi>0}$: convergence accuracy;
\STATE {\textbf{Output:}} Optimal solution, i.e., ${\bm \nu}_{n}^{*}$;
\STATE {\textbf {Solve:}} Objective function, i.e., $f(\bm \nu_{n}^{(0)})$;
\STATE {\textbf {repeat}}
\STATE \hspace{0.2cm} \textbf {for} ${r = {r+1}}$ {\rm \textbf{do}} 

\STATE \hspace{0.5cm} \textbf {for} ${n = {1}}$ to ${N}$ {\rm \textbf{do}}

{\STATE \hspace{1cm}1. Solve~(\ref{eq:37}) to calculate optimal ${\bm \nu}_{n}^{*} \ \{\forall n\in N\}$, for fixed 

\STATE \hspace{1.2cm} $\nu_{l}$ $\{\forall l \in N\}, l\neq n$;}

\STATE \hspace{1cm}2. Obtain ${\bm \nu}_{n}^{*}=\{{\bm \nu}_{1}^{*},\dots,{\bm \nu}_{l}^{*},\dots,{\bm \nu}_{N}^{*}\}$

\STATE \hspace{1cm}3. Set ${\bm \nu}_{n}^{(r+1)}={\bm \nu}_{n}^{*}$ and calculate $f{(\bm \nu_{n}^{(r+1)}})$;

\STATE \hspace{1cm}4. Perform iterations till stopping criteria meets i.e., 
\STATE \hspace{1.2cm}
$\frac{{{||f({\nu}_{n}^{(r+1)}}})-{{f({\nu}_{n}^{(r)}}})||}{{{f({\nu}_{n}^{(r+1)}}})}<{\xi}$;
\STATE \hspace{0.7cm} \textbf {end}
\STATE \hspace{0.2cm} \textbf {end}

\STATE \textbf{until} convergence;
\end{algorithmic}
\end{algorithm}
\subsection{Transmit BF Optimization}
 \label{sec:IV-B}
By exploiting phase shifts obtained from Algorithm~\ref{alg:alg1}, the effective channels of ${\mathtt {U}^{net}}$ users can be modelled easily. Further, at the ${\rm {AP}_\mathtt {U}}$ without phase shifts consideration, the transmit BF is performed for the given channel models by solving the following sub-problem
\setlength{\abovedisplayskip}{2pt}
\setlength{\belowdisplayskip}{2pt}
  \begin{subequations}
        \begin{empheq}[left=]{align}
            &({\rm P_{2.4}}): \underset{\textbf{W}_{\mathtt U}}{\min}
            ~~\hspace{5pt}{{\displaystyle\sum_{k=1}^{K_{\rm I}}}}||{\textbf{w}}_{{\rm R}_{k}}||^{2}
              \label{eq:38a}\\
             & \hspace{35pt} \text{s.t.} \hspace{14pt} \Gamma_{{\rm R}_{k}} \geq {\Gamma _{{\rm R}_{k,{\rm min}}}}, \ \forall k \in {K_{\rm I}},\label{eq:38b}
        \end{empheq}
            \end{subequations}
$({\rm P_{2.4}})$ is similar to $({\rm P_{2}})$ without considering IRS phase shifts. The problem can be solved feasibly using Heuristic approach discussed in Section~\ref{sec:III-A}.
\subsection{Computational Complexity Analysis of Alternative Solution}
 \label{sec:IV-C}
Algorithm~\ref{alg:alg3} provides an overall summary of the alternative algorithm for solving $(\rm P_{2})$ optimization problem.The low-complexity alternative solution provides closed-form expressions for optimization variables, that are computationally efficient to solve. Furthermore, by decoupling the variables in $({\rm P_{2.3}})$ and $({\rm P_{2.4}})$ and solving them separately, the complexity of solving phase shifts can be reduced significantly. The complexity of solving $({\rm P_{2.3}})$ is only $\mathcal{O}\big(I_{0}N\big)$, where $I_{0}$ shows the number of iterations required to achieve convergence. The complexity of solving $({\rm P_{2.4}})$ via MMSE approach is  $\mathcal{O}\big(K_{\rm I}M^{3}\big)$. Hence, the overall complexity of solving $({\rm P_{2.3}})$ and $({\rm P_{2.4}})$ can be shown to be $\mathcal{O}\big(I_{out}\big(I_{\lambda}\big(K_{\rm I}M^{3}\big)+I_{inn}\big(N\big)\big)$, where $I_{inn}$ and $I_{out}$ show inner and outer iterations to reach convergence, respectively. 
\begin{algorithm}[!t]
\small
\caption{Overall Low Complexity Alternative Algorithm for Solving $(\rm P_{2})$}\label{alg:alg3}
\begin{algorithmic}
\STATE 
\STATE {\textsc{\textbf{Initialization}:}} $r$: the iteration number, ${\xi>0}$: convergence accuracy, $\textbf w_{{\rm R}_{k}}^{(0)}$, ${\bm \nu}^{(0)}$: initial feasible solutions;

\STATE {\textbf {repeat}}

{\STATE \hspace{0.5cm}}1. Update IRS phase shifts by solving $({\rm P_{2.3}})$.

{\STATE \hspace{0.5cm}}2. Update transmit beamformers by solving $({\rm P_{2.4}})$.

\STATE \hspace{0.5cm}3. Perform iterations until the objective value of $(\rm P_2)$
{\STATE \hspace{0.7cm}}  falls below threshold accuracy i.e.,~(\ref{eq:15a})$\leq \xi$.

\STATE \textbf{until} convergence;
\end{algorithmic}
\end{algorithm}
\section{EE Optimization at ${\mathtt {I}^{net}}$}
\label{sec:V}
The problem of phase shifts optimization is solved in Sections~\ref{sec:III-B} and~\ref{sec:IV-A} by exploiting the SDP approach using SDR and an iterative solution via EBCD methods, respectively. By exploiting these phase shifts of ${\mathtt {U}^{net}}$ using the phase cooperation approach, the problem of EE maximization at ${\mathtt {I}^{net}}$ can be feasibly solved.  
In this section, we first optimize the EH coefficient ${\varphi_k}$ of ${\mathtt {I}^{net}}$ by deriving a closed-form solution that can effectively improve the resultant EE. For a given BF vector and by employing phase shifts of ${\mathtt {U}^{net}}$, the optimization problem $({\rm P_1})$ will become
    \begin{subequations}
        \begin{empheq}[left=]{align}
            &({\rm P_{1.1}}): \underset{\varphi}{\max}\hspace{12pt}\frac {{\mathcal R_{k}^{{\mathtt {I}^{net}}}}}{1/\eta{{\sum_{k=1}^{K_{\rm E+I}}}}||{\textbf{w}}_{{\rm D}_{k}}||^{2}+P_C}\label{eq:39a}
            \\
            & \hspace{35pt} \text{s.t.} \hspace{14pt} 
            ~(\ref{eq:14b}),~(\ref{eq:14d}),~(\ref{eq:14g}) 
        \end{empheq}
    \end{subequations} 
To solve $({\rm P_{1.1}})$ a closed-form solution for ${\varphi_k}$ is derived by using the achievable data rate constraint of ${\mathtt {I}^{net}}$ given as
\begin{align}
     {\mathcal R}_{{\rm D}_{k}} >  {\mathcal R}_{{\rm D}_{k,{\rm min}}}, \ \ \forall k \in K_{\rm E+I},
     \label{eq:40}
\end{align}
where ${\mathcal R}_{{\rm D}_{k,{\rm min}}}$ is the minimum predefined data rate. Since, ${{\rm D}_{k}}$ can only decode data when~(\ref{eq:40}) is met. Consequently, the EH coefficient obtained by solving~(\ref{eq:40}) is the optimize ${\varphi_k}$, and it is given by
\begin{align}
     {{\varphi_k}}=
     1-{\frac{{\Gamma _{{\rm D}_{k,{\rm min}}}}\sigma_{k}^2}{{{|\textbf{f}_{{{\rm D}_{k}}}^H\textbf{w}_{{\rm D}_{k}}|^2}}-{\Gamma _{{\rm D}_{k,{\rm min}}}}{\displaystyle\sum_{j\neq k}{|\textbf{f}_{{{\rm D}_{k}}}^H\textbf{w}_{{\rm D}_{j}}|^2}}}}-\epsilon.
     \label{eq:41}
\end{align}
\\
\textit{Proof}: Please, refer to the Appendix.
\par
By defining a pre-threshold ${\bar E_{\mathcal H_k}}$ and using~(\ref{eq:41}), the constraint~(\ref{eq:14b}) can be easily satisfied. Next, by substituting optimize ${\varphi_k}$ in $({\rm P_1})$, the resultant problem can be simplified to a transmit BF optimization similar to $({\rm P_{2.1}^{*}})$, which can be feasibly solved by using Sec~\ref{sec:III-A} heuristic approach.

\section{Simulation Results and Discussion}
\label{sec:VI}
\begin{figure}
\centering
\includegraphics[width=2.8in,height=1.5in]{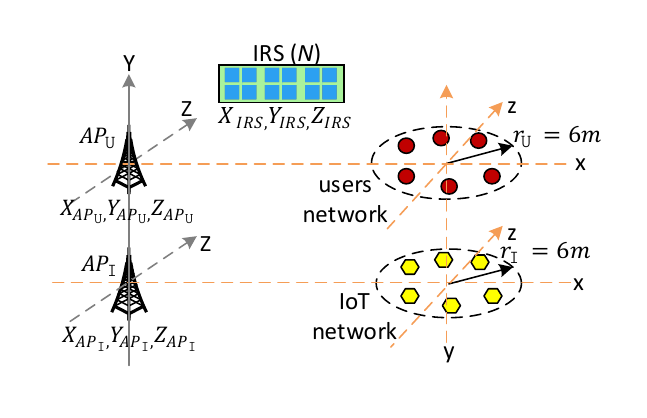}\caption{\small {3D-setup for network deployment.}}
\label{fig:2}
\end{figure}

\setlength\belowcaptionskip{-3ex}
\setlength\abovecaptionskip{-2ex}
In this section, we present numerical results to evaluate the performance of the proposed framework and optimization schemes. The performance is assessed by considering EE maximization for both users and PS-SWIPT based SS-IoT networks. We consider two APs; ${\rm {AP}_\mathtt {U}}$ and ${\rm {AP}_\mathtt {I}}$, equipped with multiple antennas for serving ${\mathtt {U}^{net}}$ and ${\mathtt {I}^{net}}$, whereas within each network we consider 6 users and 6 SS-IoT devices
equipped with single antennas and distributed randomly in a circular region centred at $({10,0})$ with radius $r_{\{\texttt{U,I}\}}=6\ {\rm m}$ separately. The APs are located at $(0;0;0)$ and the IRS is deployed at a location of $(X_{\rm IRS}=6\ {\rm m};Y_{\rm IRS}=8\ {\rm m}; 0)$. For further simulation setup, the 3-D coordinate system with network deployment is shown in Fig.~\ref{fig:2}. The channel links between the APs-IRS and IRS to ${\mathtt {U}^{net}}$ and ${\mathtt {I}^{net}}$ are mutually independent; hence, we adopt a line-of-sight (LoS) channel model, i.e., the Rician fading model for APs-IRS, whereas the channel links between the IRS to ${\mathtt {U}^{net}}$ and ${\mathtt {I}^{net}}$ follow Rayleigh distribution as the users and devices are deployed randomly. For the considered framework, the channel model is expressed as
\begin{align}
    \textbf{G}_{\rm R,D}&={\sqrt {P(L_x)}}\Bigg \{{\sqrt{\frac{\kappa}{1+\kappa}}} \textbf{\~G}{^{\rm LoS}}+{\sqrt{\frac{\kappa}{1+\kappa}}}\textbf{\~G}{^{\rm NLoS}}\Bigg \},
\label{eq:42}  \\
    {\textbf h}_{\{{\rm R,D}\}_{{\rm r},k}}^{H}&={\sqrt {P(L_x)}}\Bigg \{{\sqrt{\frac{\kappa}{1+\kappa}}} \textbf{\~h}_{{\{\rm R,D\}}_{{\rm r},k}}^{H (\rm LoS)}+{\sqrt{\frac{1}{1+\kappa}}}\textbf{\~h}_{\{{\rm R,D}\}_{{\rm r},k}}^{H (\rm NLoS)}\Bigg \},
\label{eq:43}  \\
\textbf{g}_{{{\{\rm R,\rm D\}}_{{\rm d},k}}}^H&={\sqrt{{P(L_x)}}}\textbf{\~g}_{{{\{\rm R,\rm D\}}_{{\rm d},k}}}^H.
\label{eq:92}  
\end{align}
Here,  ${P(L_x)}$ is the distance dependent path-loss model given by ${P(L_x)}=C_0\big({{d_{x}}/{d_{0}}}\big)^{-a_{x}}$,
where $C_{0}$ is the the path loss at the reference distance $d_{0}$, $d_{x}$, and $a_{x}$ $\forall x \in {\{\rm APs-IRS},\ {\rm IRS}-{{{\rm R}_k}},{{{\rm D}_k}}, \ {\rm APs}-{{{\rm R}_k}},{{{\rm D}_k}}\}$ shows the distances and path loss exponents of the individual links, respectively. 
Whereas, $a_{\rm APs-IRS}=2$, $a_{\rm IRS-{{\rm R}_k},{{\rm D}_k}}=2.5$, and $a_{\rm APs-{{\rm R}_k},{{\rm D}_k}}=3.5$ show their respective path losses. For the proposed network model under consideration, we consider a bandwidth of 1 MHz and noise variance $\sigma^2_{k}$, $\forall k \in \{\mathtt {U,I}\}$. 
Moreover, we consider $\kappa=5\ {\rm dB}$ as the Rician constant factor, $\textbf{{\~G}}^{\rm LoS}$, $\textbf{\~h}_{{\{\rm R,D\}}_{{\rm r},k}}^{H (\rm LoS)}$, and $\textbf{{\~G}}^{\rm NLoS}$, $\textbf{\~h}_{{\{\rm R,D\}}_{{\rm r},k}}^{H (\rm NLoS)}$ represent the LoS and non-LoS (NLoS) components of reflected paths, respectively. The elements of $\textbf{{\~G}}^{\rm NLoS}$ and $\textbf{\~h}_{{\{\rm R,D\}}_{{\rm r},k}}^{H (\rm NLoS)}$ follow the Rayleigh fading model with identical independent distribution (i.i.d.). The channel coefficients for the direct path, i.e., without IRS scenario $\textbf{g}_{{{\{\rm R,\rm D\}}_{{\rm d},k}}}^H$ are generated at random from a Gaussian normal distribution. Furthermore, we consider that all users and IoT devices have the same SINR requirements, i.e., $\Gamma_{{\{\rm R,D}\}_{k,\rm min}}=4$ dB, $\forall k$. Some other parameters we consider are $\eta_{{\{\rm R,D}\}_{k}}=0.8$, $P_{C_{{\{\mathtt {U,I}}\}}}=5$ dBm, $\mu_{{\rm D}_{k}}=0.8$, and $\epsilon_{{\rm D}_{k}}=10^{-5}$.
\par
The simulation results for the proposed framework is evaluated based on following baseline schemes.
\begin{itemize}
    \item 
{\textbf {Baseline Scheme 1 (AO with IRS):}} The maximization problem is solved for ${\mathtt {U}^{net}}$ using transmit BF and phase shifts optimization by applying AO with IRS. The BF vectors are optimized using heuristic approach whereas the phase shifts are optimized via computationally efficient SDR scheme. The CVX tool~\cite{grant2014cvx} is used to solve the phase optimization, with threshold accuracy and line-search accuracy set to $\xi_{\{{\mathtt {U,I}}\}}=0.001$ and $\zeta_{\{{\mathtt {U,I}}\}}=0.1$, respectively. For the proposed approach, we consider $10000$ Gaussian randomizations. The EE maximization is then performed for ${\mathtt {I}^{net}}$ by using the heuristic BF algorithm and the ${\mathtt {U}^{net}}$ optimal phase shifts $\Theta$.
\item
{\textbf {Baseline Scheme 2 {(Low Complexity Alternative Solution with IRS (LCAS)):}}} The phase shifts and BF vectors are optimized via low complexity alternative solution. For IRS phase optimization, EBCD approach with $3000$ iterations and a convergence accuracy of ${\xi=0.001}$ is used.
\item
{\textbf {Baseline Scheme 3 {(Discrete IRS Phase Shifts (DPS)):}}} A discrete set of IRS phase shifts is considered by using the approaches of~\cite{hua2020intelligent} and~\cite{9559408}. The optimal discrete phases are obtained as ${\nu}_{n}^{*}=exp(j\beta_{n,{\mathscr l^*}})$ $ \forall n\in \{1,N\}, \forall \mathscr l^* \in \{1,\mathcal L\}$, where $\mathcal L=2^{\mathcal B_0}$ shows the total number of discrete phase shifts, and $\mathcal B_0$ represent the IRS phase resolution bits. The optimal set of DPS are selected from SDR or EBCD approach for employing phase optimization. 
\item {\textbf {Baseline Scheme 4 {(Random IRS phase shifts (RPS)):}}} The phase shifts of IRS are randomly generated, and the BF optimization is performed by heuristic schemes. For SWIPT network PS ratio is obtained via closed-form expression.
\item {\textbf {Baseline Scheme 5 {(Without IRS (W/O IRS)):}}} The system model is reduced to conventional wireless PS-SWIPT based ${\mathtt {I}^{net}}$ without IRS support. The APs employ a direct communication link in the presence of obstacles, considering the practical path loss channel model.
\end{itemize}
\par 
To benchmark the performance of MMSE scheme, we also analyze rate curves of above baseline schemes using ZFBF method discussed in Section III-A.
\subsection{Convergence Analysis of Algorithm 1 and 3}
\begin{figure}[!t]
\centering
\includegraphics[width=3.1in,height=2.1in]{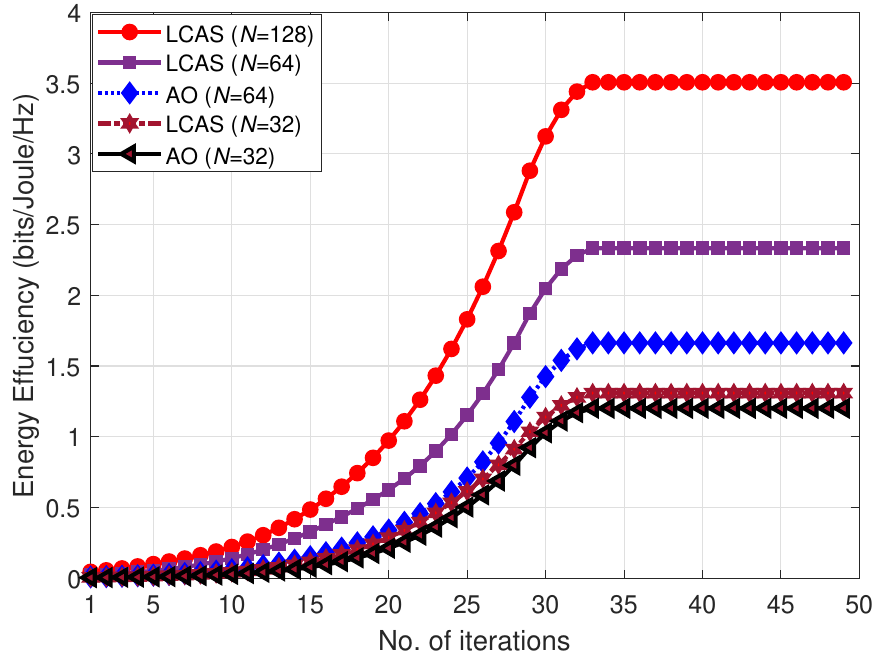}
\caption{\centering {{Convergence behaviour of Algorithms~\ref{alg:alg1} and~\ref{alg:alg3}.}}}
\label{fig:3}
\end{figure}

First, we analyze the convergence behaviour of algorithms~\ref{alg:alg1} and~\ref{alg:alg3} with continuous phase shifts as presented in Fig.~\ref{fig:3}. We analyze the EE performance curves versus iteration numbers for the discussed framework using a fixed number of APs antennas and varying IRS reflection elements, i.e., $M_{\mathtt U}=10$ and $N = 32,64,$ and $128$, respectively. The performance curves in Fig.~\ref{fig:3} prove the convergence of the AO and the LCAS algorithms, clearly indicating that after a few iterations, the EE curve shows stable convergence behaviour at a single point, thus confirming the convergence of both algorithms. It is also noteworthy that when the algorithms converge, they achieve a relatively similar performance at a small number of IRS elements, validating the theoretical and computational analysis discussed in Sections~\ref{sec:III-B} and~\ref{sec:IV-A} for Algorithms~\ref{alg:alg1} and~\ref{alg:alg3}.
\setlength\belowcaptionskip{-4ex}
\setlength\abovecaptionskip{1ex}
\subsection{Performance Analysis of ${\mathtt {U}^{net}}$}
\begin{figure}
\centering
\includegraphics[width=3in]{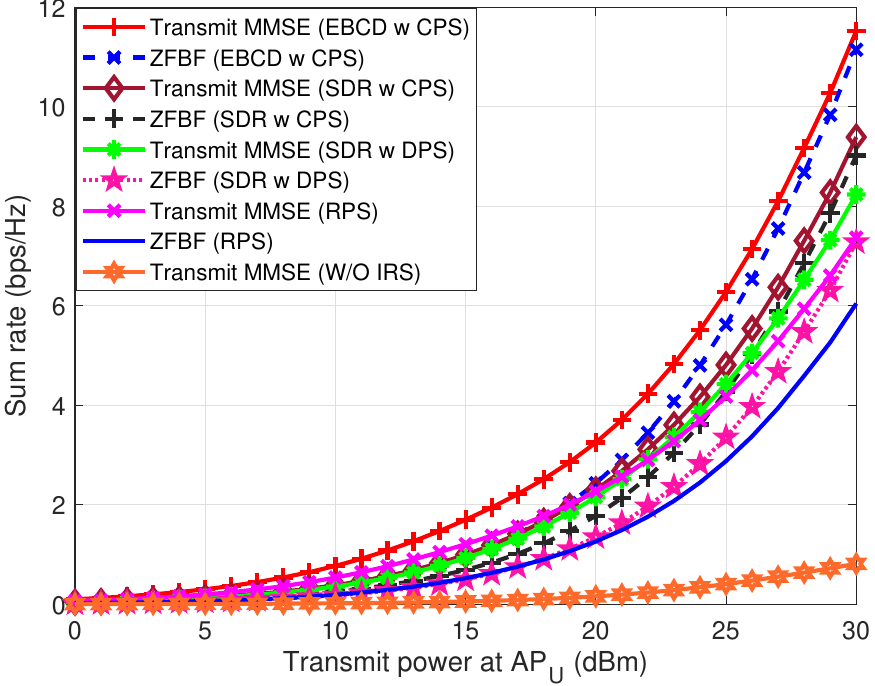}
\caption{\centering {Sum rate versus $P_{\rm AP_{\mathtt {U}}}$.}}
\label{fig:4}
\end{figure}

\begin{figure}
\centering
\includegraphics[width=3in]{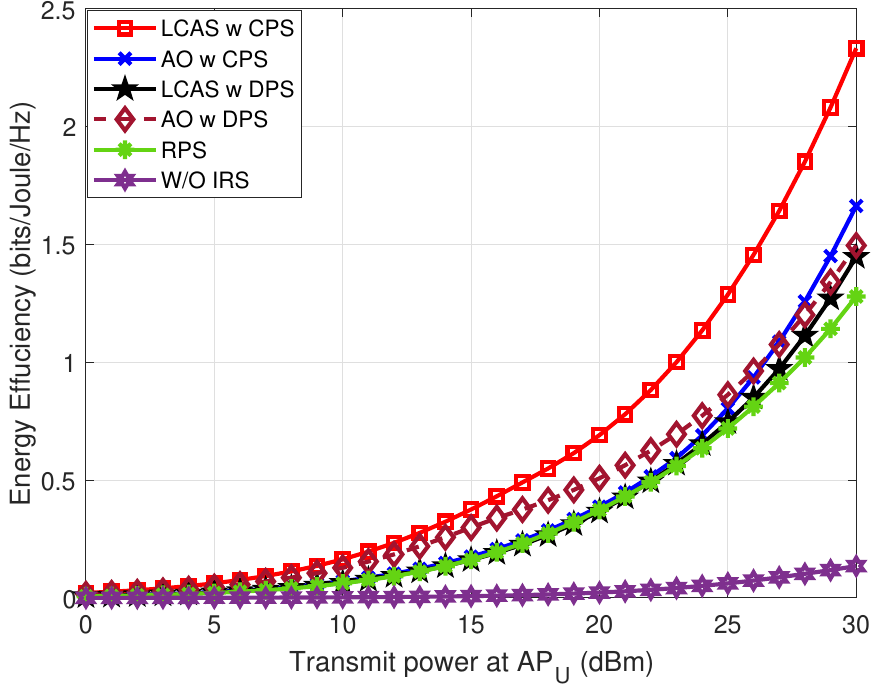}
\caption{\centering {Energy efficiency versus $P_{\rm AP_{\mathtt {U}}}$.}}
\label{fig:5}
\end{figure}

For the proposed ${\rm PC}_{Net}$, the optimal phase shifts of ${\mathtt {U}^{net}}$ are exploited by ${\mathtt {I}^{net}}$ to improve the EE performance without constraining resources. Hence, we first investigate the performance of ${\mathtt {U}^{net}}$ from some aspects with the outlined baseline schemes. The sum rate and EE performance metrics are plotted against the maximum transmit power, i.e., $P_{\rm AP_{\mathtt {U}}}$ at ${\rm {AP}_\mathtt {U}}$, respectively, in figures~\ref{fig:4} and~\ref{fig:5}. As expected, with increasing $P_{\rm AP_{\mathtt {U}}}$, both the sum rate and EE curves show an increasing trend. By employing Algorithms~\ref{alg:alg1} and~\ref{alg:alg3}, the curve achieves a significant performance gain compared to other baseline schemes. Furthermore, it is noted that the proposed scheme with continuous phase shifts (CPS) and DPS performs better than RPS and W/O IRS approaches, which confirms that IRS can improve the performance of the WIT networks. For transmit BF comparison, it is shown from gain curves that ZFBF is asymptotically ideal at high values of transmit power, whereas transmit MMSE BF achieves higher performance than ZFBF through the entire range of transmit power; however, MMSE performance relies strongly on careful adjustment of the controlling parameter $\lambda_{k}$.

\setlength\belowcaptionskip{-3ex}
\setlength\abovecaptionskip{2ex}
\par
\begin{figure}[!t]
\centering
\includegraphics[width=3in]{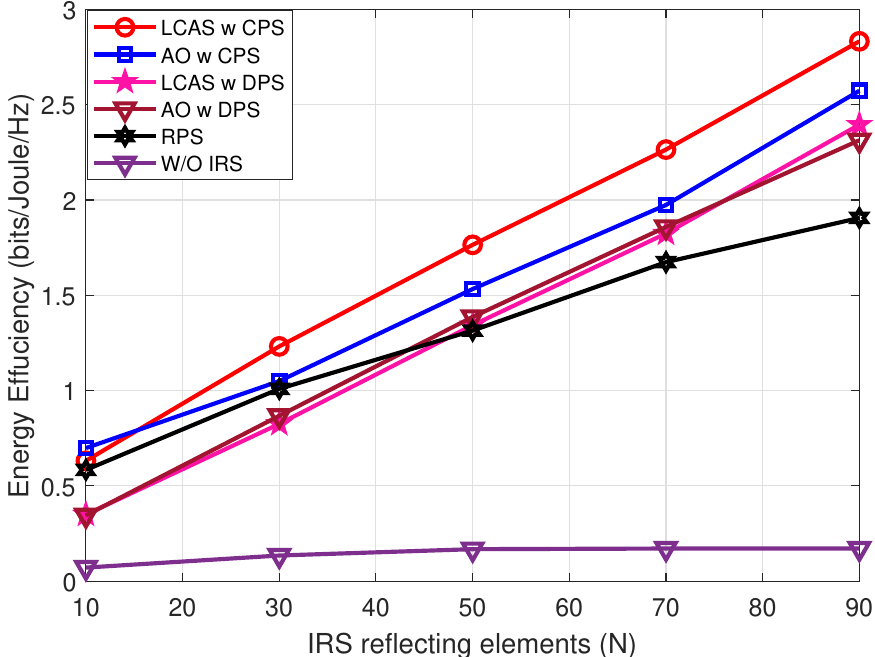}
\caption{\centering {Energy efficiency versus reflecting elements $N$.}}
\label{fig:6}
\end{figure}

Next, we investigate the EE performance vs. number of IRS reflection elements $N$, which can be easily observed to have a monotonically increasing trend as shown in Fig.~\ref{fig:6}. With increasing $N$, the respective energy and information signals become stronger at the intended receiver end; hence, the EE performance of all baseline schemes shows improvement compared to the W/O IRS case, and the performance of proposed Algorithms~\ref{alg:alg1} and~\ref{alg:alg3} outperforms other schemes. Moreover, even the EE curve when employing the RPS scheme also shows higher EE performance than the W/O IRS scheme, particularly with large $N$, implying that the IRS can bring a substantial rise in the performance gain of conventional networks. Moreover, it supports the utilization of IRS in place of active antennas and expensive RF modules to reduce system costs.
\par
\begin{figure*}[!t]
\captionsetup[subfigure]{font=small,labelfont=small}
\centering
\hskip -3.5ex
\subfloat[\scriptsize ] {\includegraphics[width=2.4in,height=2in]{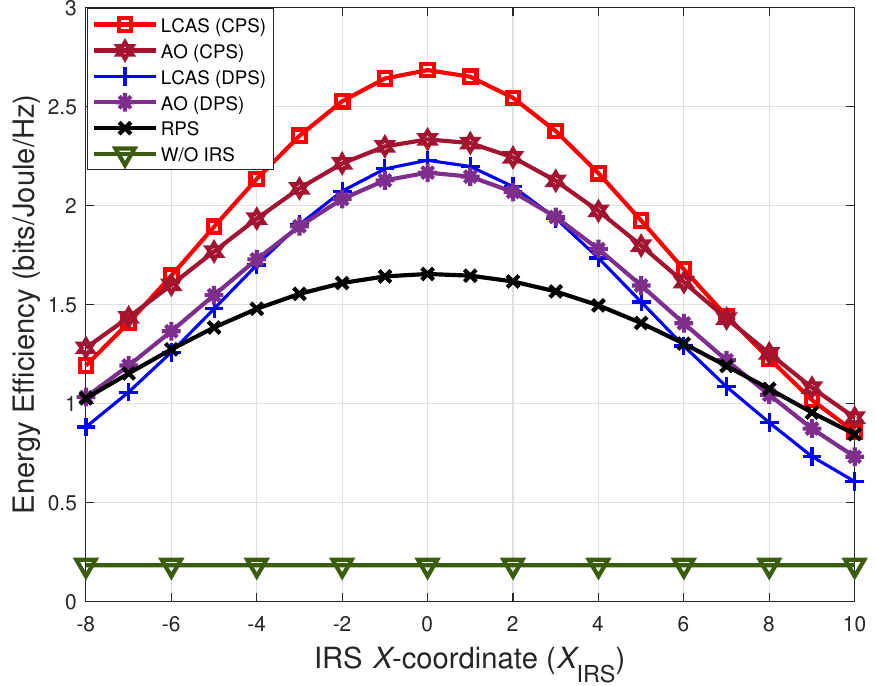}%
\label{fig:7(a)}}
\hfil
\hskip -1ex
\subfloat[\scriptsize ] {\includegraphics[width=2.4in,height=2in]{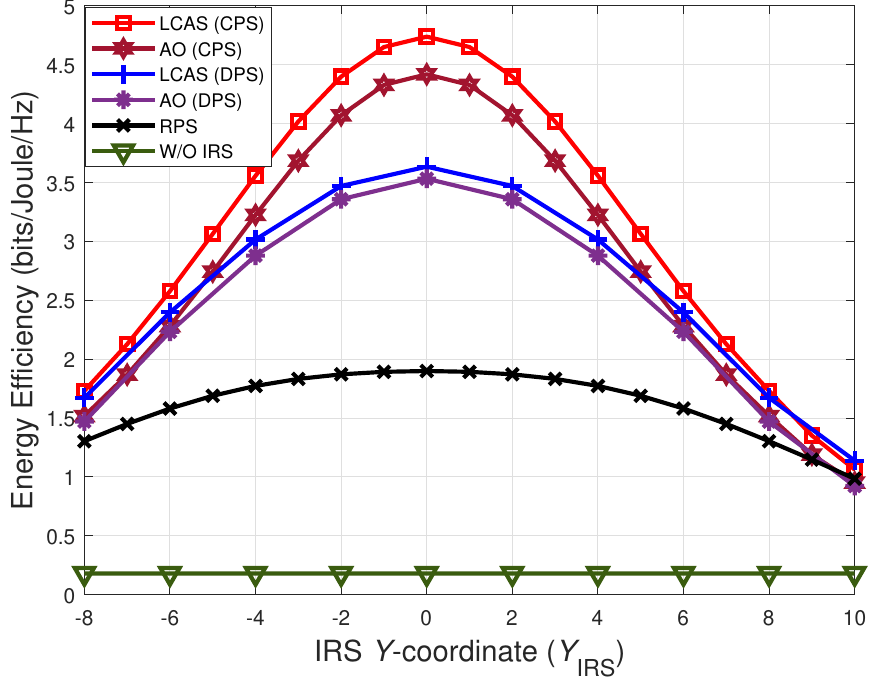}%
\label{fig:7(b)}}
\hfil
\hskip -1ex
\subfloat[\scriptsize ] {\includegraphics[width=2.4in,height=2in]{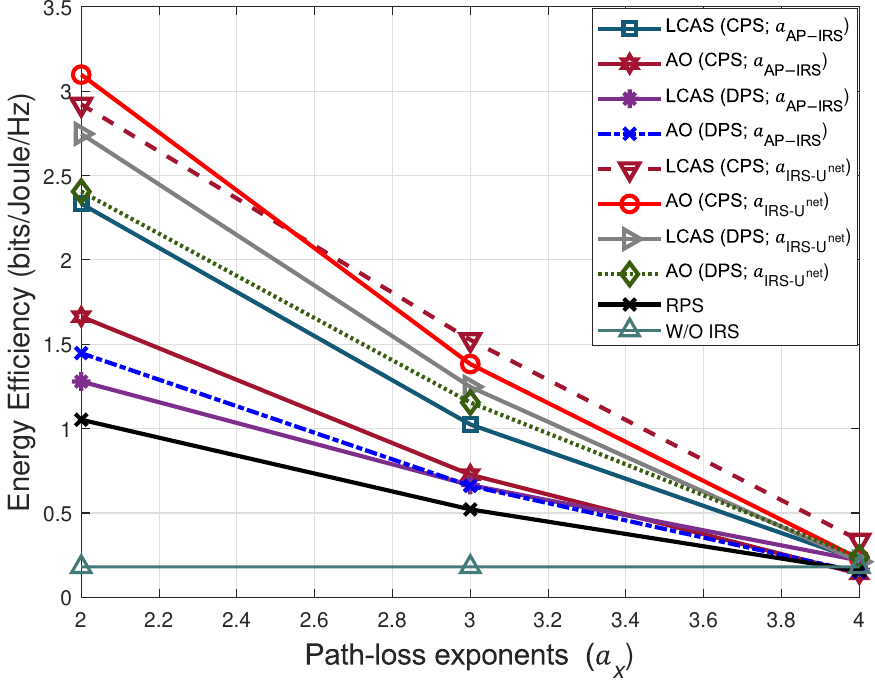}%
\label{fig:7(c)}}
\caption{\small {Energy efficiency performance w.r.t IRS deployment; (a) EE vs IRS $X$-coordinate $(X_{\rm IRS})$, (b) EE vs IRS $Y$-coordinate $(Y_{\rm IRS})$, and (c) EE versus path-loss exponents $(a_x, \ \forall x \in {\{\rm AP-IRS},\ {\rm IRS}-{{{\rm R}_k}}\})$. \\}}
\label{fig:7}
\end{figure*}

 
We also study the impact of IRS deployment to enhance the corresponding signal and energy reflections at the intended receiver side, which in turn improves EE performance, as shown in Fig.~\ref{fig:7}. To investigate this, we compare the EE and IRS $X$ and $Y$-coordinates; $(X_{IRS},Y_{IRS})$ varying from $-8$ to $10$ in figures~\ref{fig:7(a)} and~\ref{fig:7(b)}, respectively. It can be seen from the figures curves that EE first rises as the IRS's $X$ and $Y$-coordinates rise, and then it starts to decline. The fact is that increasing distance between AP and IRS makes large-scale fading more prominent as the path loss exponent increases, causing lower energy reception and reducing the primary benefit of IRS deployment in conventional networks. To demonstrate the concept further, we analyse the EE with path-loss exponents associated with the two-fold IRS reflecting paths, i.e., AP-IRS and ${\rm IRS}-{{{\rm R}_k}}$, as expressed in Fig.~\ref{fig:7(c)}. The figure clearly illustrates that the EE decreases with respect to path-loss exponents owing to increasing large-scale fading, which would lower energy receipt and minimize the benefits of IRS deployment in a proposed network. Thus, for improving a network's performance, the optimal deployment of IRS plays a vital role.
\subsection{Performance Analysis of ${\mathtt {I}^{net}}$}
\begin{figure}
\centering
\includegraphics[width=3in]{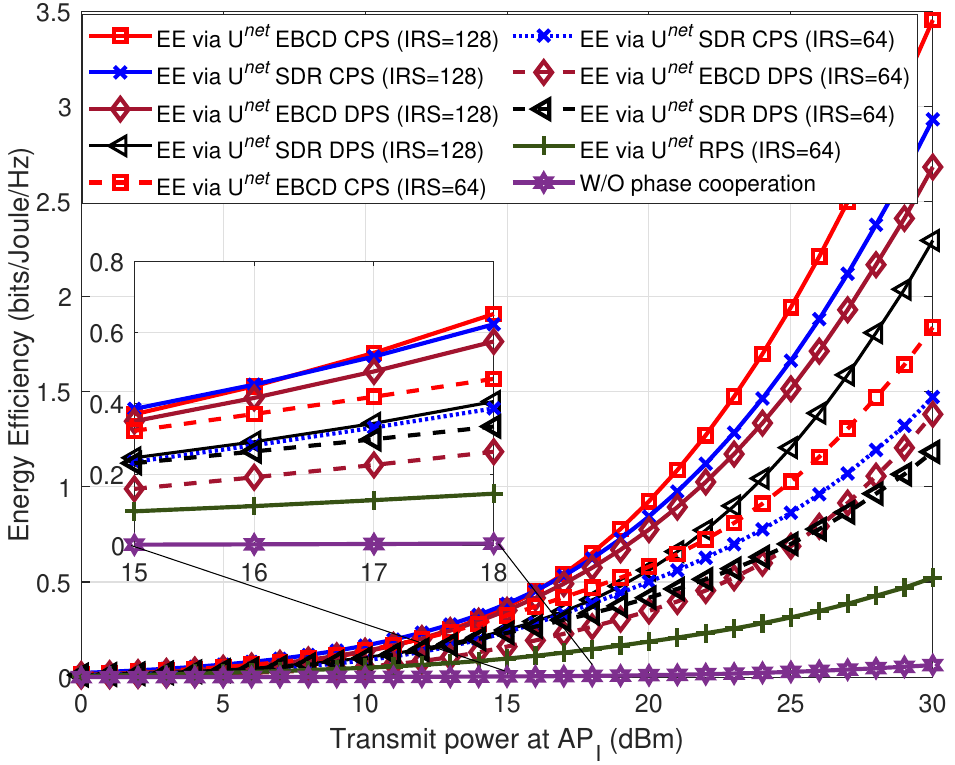}
\caption{\centering {Energy efficiency versus $P_{\rm AP_{\mathtt {I}}}$.}}
\label{fig:8}
\end{figure}

In this section, by exploiting the optimal phase shifts of ${\mathtt {U}^{net}}$ obtained via SDR and EBCD approaches, the EE performance is evaluated for a PS-SWIPT based SS-IoT network via effective phase cooperation. For the given optimal phase shifts and optimal PS coefficients, the EE curves vs maximum transmit power available at $P_{\rm AP_{\mathtt {I}}}$ is shown in Fig.~\ref{fig:8} for the considered baseline schemes. As expected from ${\mathtt {U}^{net}}$ results, the EE of ${\mathtt {I}^{net}}$ also rises with an increasing $P_{\rm AP_{\mathtt {I}}}$ trend. By employing a heuristic BF approach for the given optimal $\Theta$ and $\varphi_k$, the EE curves of ${\mathtt {I}^{net}}$ achieve a significant performance gain compared with the W/O IRS scheme, which confirms that using IRS phase cooperation we can enhance SWIPT IoT networks performance.
\par
Next, we unveil the potential of the varying IRS elements of ${\mathtt {U}^{net}}$ on the EE performance of ${\mathtt {I}^{net}}$, also in Fig.~\ref{fig:8}. We consider different IRS reflecting elements for optimizing ${\mathtt {U}^{net}}$ $\Theta$ and exploiting these constant optimal phase shifts via phase cooperation; EE curves are evaluated for ${\mathtt {I}^{net}}$. The performance curves validate the effectiveness of phase cooperation with a monotonically increasing trend, as shown in Fig.~\ref{fig:8} results. Again, the EE achieved by the baseline scheme increases compared to the W/O IRS phase cooperation case. This implies that with improved phase shifts, a substantial rise in performance gains for PS-SWIPT based SS-IoT networks can be achieved.
\begin{figure}
\centering
\includegraphics[width=3in]{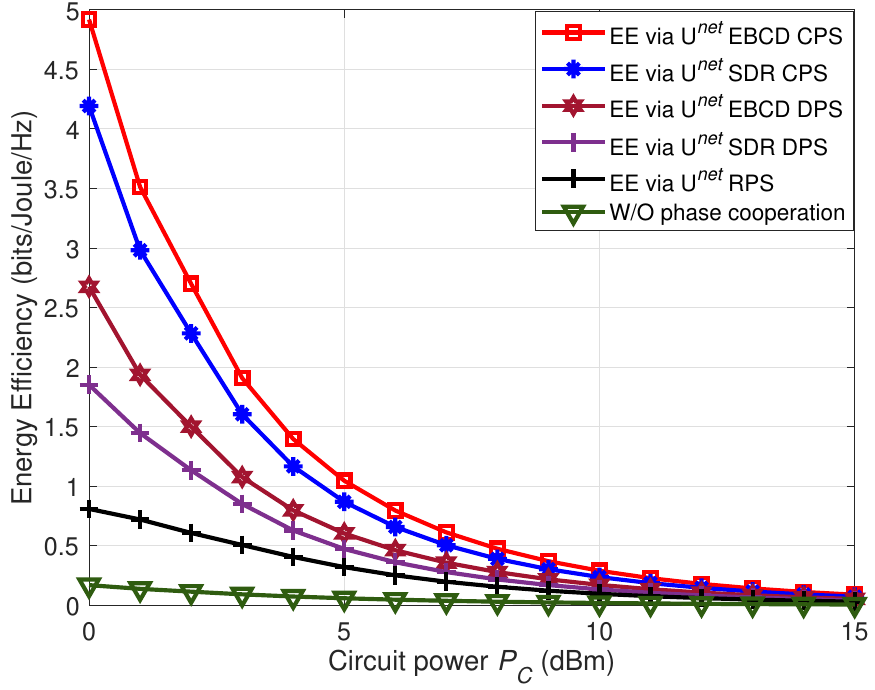}
\caption{\centering {Energy efficiency versus $P_{C}$ at ${\mathtt {I}^{net}}$.}}
\label{fig:9}
\end{figure}
\par
Fig.~\ref{fig:9} shows the PS-SWIPT network EE performance against the overall circuit power consumption $P_{C}$ at ${\rm {AP}_\mathtt {I}}$. We set transmit antennas at $M_\mathtt {I} = 10$, exploit phase shifts with $N = 64$, and $P_C$ ranges from $0$ to $15$ dBm. Further, we considered a different transmit power budget for the proposed scheme. Fig.~\ref{fig:9} curves demonstrate that the EE declines when the circuit power increases, which is consistent with the EE maximization principle. However, as $P_{C}$ further increases, the slope of the declining curve decreases since optimization is more prevalent in increasing network EE when $P_{C}$ is low. The performance of EE curves with phase cooperation outperforms those without phase cooperation.
\begin{figure}[!t]
\centering
\includegraphics[width=3in]{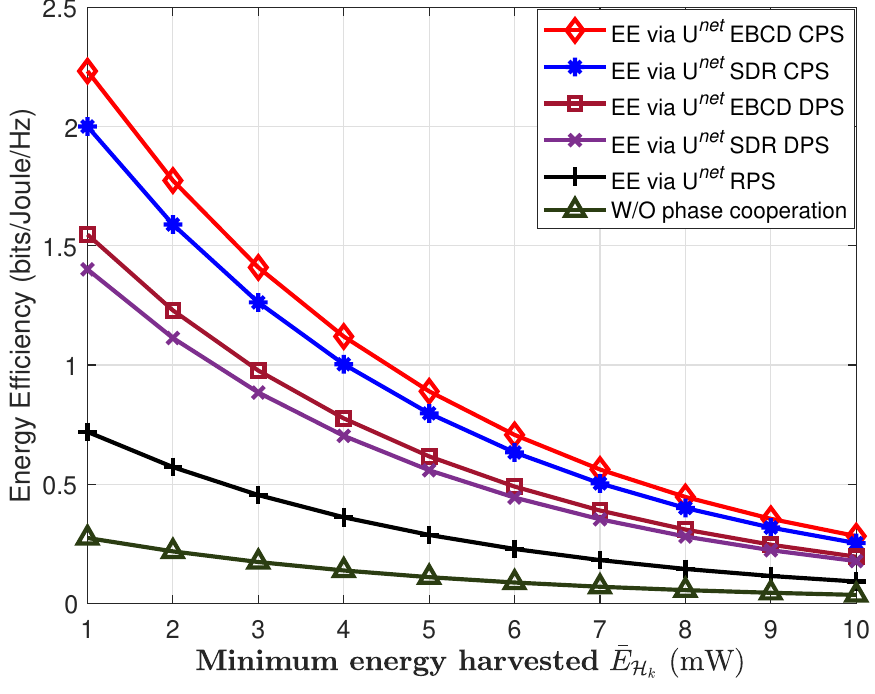}
\caption{\centering {Energy efficiency versus minimum energy harvested by ${{{\rm D}_k}}$ for information decoding.}}
\label{fig:10}
\end{figure}
\par
Finally we illustrate the EE versus minimum energy harvested requirement of SS-IoT devices, i.e., ${\bar E_{\mathcal H_k}}$ using the baseline schemes in Fig.~\ref{fig:10}. As we can see, the proposed solution with phase cooperation outperforms the benchmarks significantly. Besides, IRS-enabled phase cooperative systems are more resistant to minimum energy harvested than conventional systems, as demonstrated by the difference between the scheme with optimal phase shifts and the scheme without phase shifts with increasing minimum harvested energy.
\section{Conclusions and Future Work}
\label{sec:VII}
This article investigated a novel IRS-enabled PS-SWIPT based SS-IoT phase cooperative framework with the formulation of the EE maximization problem for designing optimal transmission BF vectors, IRS phase shifts, and PS coefficients. To solve the non-convex optimization problem, first we applied an AO to the user network to derive transmit BF vectors using closed-form heuristic solutions and optimal phase shifts using a relaxed SDP scheme. Furthermore, we have proposed an alternative solution of lower complexity by decoupling the original optimization problem into sub-problems to obtain the optimal phase shifts through an iterative EBCD approach and transmit beamformers using a heuristic scheme. Later, the EE maximisation problem is solved for the SS-IoT network by optimizing PS coefficients and exploiting optimal phase shifts of the users network via a phase cooperation approach. A quantitative comparison of the proposed framework with the baseline schemes shows that the proposed framework can achieve high EE benefits at a low level of complexity. Further, the proposed framework can significantly reduce hardware costs by deploying IRS in conventional networks. With multiple IRSs-aided phase cooperative networks, the proposed framework can be feasibly extended in the future to a multi-user MIMO SS-IoT network scenario. Further analysis of the presented network can be done using optimization techniques under imperfect CSI.
\vspace{-1ex}
\section*{Appendix}
\subsection{Derivation of Optimized EH Coefficient}
By substituting achievable rate expression in~(\ref{eq:40}) we obtained
\begin{align}
    \ {\rm log}_2 (1+\Gamma_{{\rm D}_{k}}^{\rm ID}) \geq {\mathcal R}_{{\rm D}_{k,{\rm min}}}, \ \ \forall k \in K_{\rm E+I}.
    \label{eq:44}
\end{align}
Substituting SINR from~(\ref{eq:9}) we have
\begin{align}
      \ {\rm log}_2 \Biggl(1+ \frac{{(1-{\varphi_k})}{|(\textbf{h}_{{{\rm D}_{}\rm r},k}^H \Theta \textbf{G}_{\rm D}+\textbf{g}_{{{\rm D}_{}\rm d},k}^H)\textbf{w}_{{\rm D}_{k}}|^2}}{\displaystyle\sum_{j\neq k}{(1-{\varphi_k})}{|(\textbf{h}_{{{\rm D}_{{\rm r},k}}}^H \Theta \textbf{G}_{\rm D}+\textbf{g}_{{{\rm D}_{}\rm d},k}^H)\textbf{w}_{{\rm D}_{j}}|^2+\sigma_{k}^2}}\Bigg)\geq{\mathcal R}_{{\rm D}_{k,{\rm min}}}.
      \label{eq:45}
\end{align}
Let us substitute $\textbf{f}_{{{\rm D}_{k}}}^H={(\textbf{h}_{{{\rm D}_{{\rm r},k}}}^H \Theta \textbf{G}_{\rm D}+\textbf{g}_{{{\rm D}_{{\rm d},k}}}^H)}$ for simplification of representation
\begin{align}
     \Biggl(\frac{{(1-{\varphi_k})}{|\textbf{f}_{{{\rm D}_{k}}}^H\textbf{w}_{{\rm D}_{k}}|^2}}{\displaystyle\sum_{j\neq k}{(1-{\varphi_k})}{|\textbf{f}_{{{\rm D}_{k}}}^H\textbf{w}_{{\rm D}_{j}}|^2+\sigma_{k}^2}}\Bigg)\geq2^{{\mathcal R}_{{\rm D}_{k,{\rm min}}}}-1,
     \label{eq:46}
\end{align}
\begin{align}
     \Biggl(\frac{{(1-{\varphi_k})}{|\textbf{f}_{{{\rm D}_{k}}}^H\textbf{w}_{{\rm D}_{k}}|^2}}{\displaystyle\sum_{j\neq k}{(1-{\varphi_k})}{|\textbf{f}_{{{\rm D}_{k}}}^H\textbf{w}_{{\rm D}_{j}}|^2+\sigma_{k}^2}}\Bigg)\geq{\Gamma _{{\rm D}_{k,{\rm min}}}}, 
     \label{eq:47}
\end{align}
where ${\Gamma _{{\rm D}_{k,{\rm min}}}}$ is the minimum SINR.
\begin{align}
     {{(1-{\varphi_k})}{|\textbf{f}_{{{\rm D}_{k}}}^H\textbf{w}_{{\rm D}_{k}}|^2}}\geq
     {\Gamma _{{\rm D}_{k,{\rm min}}}}\Bigg({\displaystyle\sum_{j\neq k}{(1-{\varphi_k})}{|\textbf{f}_{{{\rm D}_{k}}}^H\textbf{w}_{{\rm D}_{j}}|^2+\sigma_{k}^2}}\Bigg),
     \label{eq:48}
\end{align}
\begin{align}
     {{(1-{\varphi_k})}{|\textbf{f}_{{{\rm D}_{k}}}^H\textbf{w}_{{\rm D}_{k}}|^2}}-{\Gamma _{{\rm D}_{k,{\rm min}}}}{\displaystyle\sum_{j\neq k}{(1-{\varphi_k})}{|\textbf{f}_{{{\rm D}_{k}}}^H\textbf{w}_{{\rm D}_{j}}|^2\geq
     +{\Gamma _{{\rm D}_{k,{\rm min}}}}\sigma_{k}^2}},
     \label{eq:49}
\end{align}
\begin{align}
     {(1-{\varphi_k})}\geq
     {\frac{{\Gamma _{{\rm D}_{k,{\rm min}}}}\sigma_{k}^2}{{{|\textbf{f}_{{{\rm D}_{k}}}^H\textbf{w}_{{\rm D}_{k}}|^2}}-{\Gamma _{{\rm D}_{k,{\rm min}}}}{\displaystyle\sum_{j\neq k}{|\textbf{f}_{{{\rm D}_{k}}}^H\textbf{w}_{{\rm D}_{j}}|^2}}}},
     \label{eq:50}
\end{align}
\begin{align}
     {{\varphi_k}}\leq
     1-{\frac{{\Gamma _{{\rm D}_{k,{\rm min}}}}\sigma_{k}^2}{{{|\textbf{f}_{{{\rm D}_{k}}}^H\textbf{w}_{{\rm D}_{k}}|^2}}-{\Gamma _{{\rm D}_{k,{\rm min}}}}{\displaystyle\sum_{j\neq k}{|\textbf{f}_{{{\rm D}_{k}}}^H\textbf{w}_{{\rm D}_{j}}|^2}}}}.
     \label{eq:51}
\end{align}
\par
To ensure that ${\varphi_k}$ always satisfy the condition mentioned in~(\ref{eq:51}), we introduce a new small slack variables $\epsilon$ such that~(\ref{eq:51}) becomes
\begin{align}
     {{\varphi_k}}=
     1-{\frac{{\Gamma _{{\rm D}_{k,{\rm min}}}}\sigma_{k}^2}{{{|\textbf{f}_{{{\rm D}_{k}}}^H\textbf{w}_{{\rm D}_{k}}|^2}}-{\Gamma _{{\rm D}_{k,{\rm min}}}}{\displaystyle\sum_{j\neq k}{|\textbf{f}_{{{\rm D}_{k}}}^H\textbf{w}_{{\rm D}_{j}}|^2}}}}-\epsilon.
     \label{eq:52}
\end{align}
\bibliographystyle{ieeetr}
\bibliography{paperbibliography}
\end{document}